\documentclass[secnumarabic,twocolumn,aps,prd,amsmath,amssymb,nofootinbib]{revtex4}
\bibpunct{}{}{,}{s}{}{,}
\usepackage{bm}

\usepackage{amsmath}
\usepackage{amsfonts}
\usepackage{amssymb}
\usepackage{graphicx}

\def\beq{\begin{equation}}
\def\eeq{\end{equation}}
\def\bea{\begin{eqnarray}}
\def\eea{\end{eqnarray}}

\def\Mp{M_{\rm pl}}
\def\d{{\rm d}}

\begin{document}

\title{Cosmological Inflation: Theory and Observations}

\author{Daniel Baumann}
\email{dbaumann@physics.harvard.edu}
\affiliation{Department of Physics, Harvard University, Cambridge, MA 02138, U.S.A.}
\affiliation{Center for Astrophysics, Harvard University, Cambridge, MA 02138, U.S.A.}

\author{Hiranya V.~Peiris}
\email{hiranya@ast.cam.ac.uk}
\affiliation{Institute of Astronomy, Cambridge University, Cambridge, CB3 0HA, U.K.}

\begin{abstract}
In this article we review the theory of cosmological inflation with a particular focus on the beautiful connection it provides between the physics of the very small and observations of the very large.
We explain how quantum mechanical fluctuations during the inflationary era become macroscopic density fluctuations which leave distinct imprints in the cosmic microwave background (CMB).
We describe the 
physics of anisotropies in the CMB temperature and polarization and discuss how CMB observations can be used to probe the primordial universe.
\bigskip

\end{abstract}

\maketitle

\tableofcontents

\newpage
\section{Introduction}

\begin{figure*}[t!]
        \includegraphics[width=.95\textwidth]{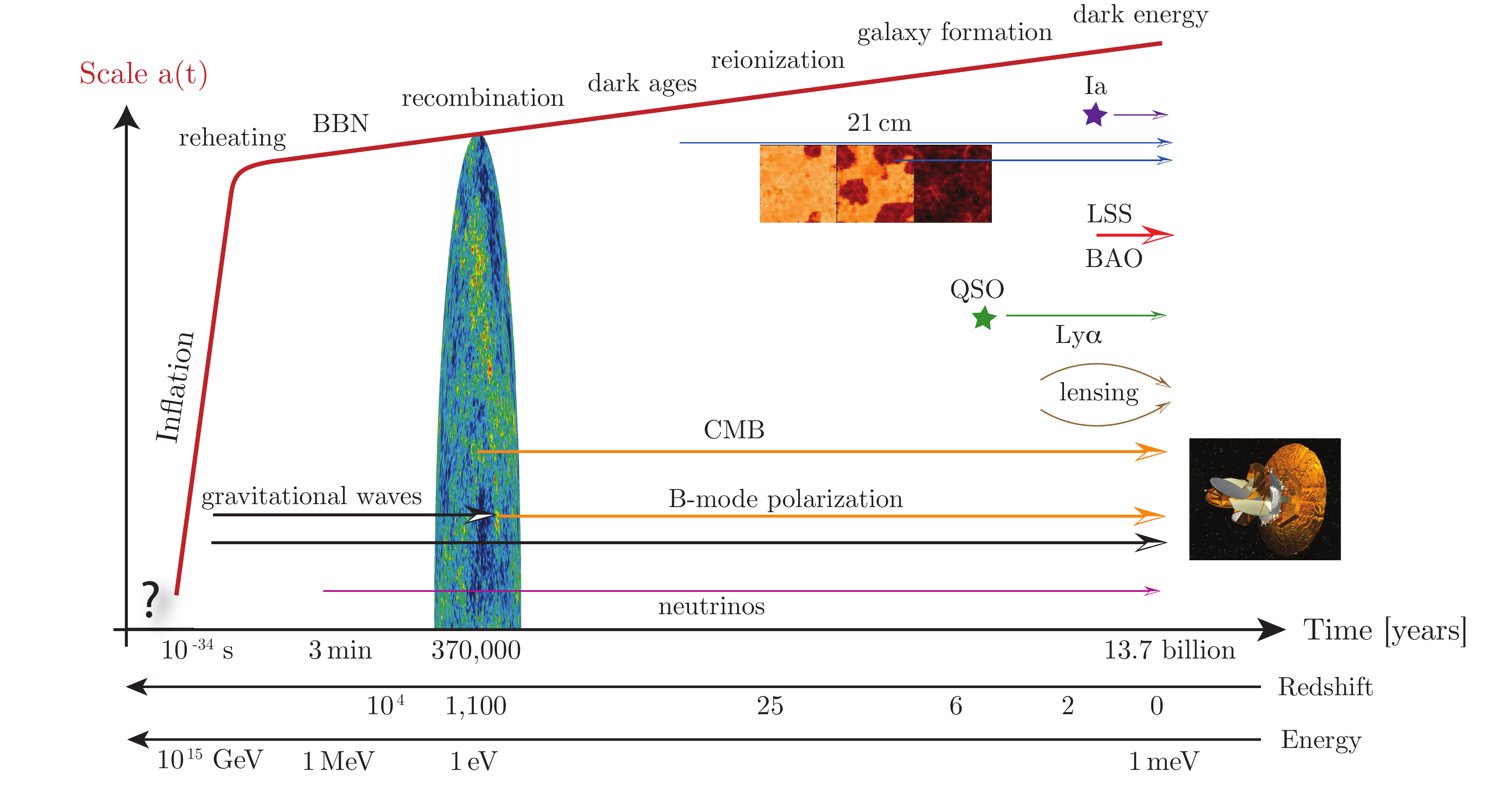}
   \caption{History of the universe. In this schematic we present key events in the history of the universe and their associated time and energy scales.  We also illustrate several cosmological probes that provide us with information about the structure and evolution of the universe. \newline
   {\it Acronyms}: BBN (Big Bang Nucleosynthesis), LSS (Large-Scale Structure), BAO (Baryon Acoustic Oscillations), QSO (Quasi-Stellar Objects; Quasars), Ly$\alpha$ (Lyman-alpha), CMB (Cosmic Microwave Background), Ia (Type Ia supernovae), 21cm (hydrogen 21cm-transition).}
    \label{fig:timeline}
\end{figure*}

The discovery of the expansion of the universe\cite{Hubble} by Edwin Hubble in 1929 heralded the dawn of observational cosmology. If we mentally rewind the expansion, we find that the universe was hotter and denser in its past.
In fact, at very early times the temperature was high enough to ionize the material that filled the universe. The universe therefore consisted of a plasma of 
nuclei, electrons and photons,  and the number density of free electrons was so high that the mean free path for the Thomson scattering of photons was extremely short. As the universe expanded, it cooled, and the mean photon energy diminished. Eventually, at a temperature of about $3000^\circ$~K, the photon energies became too low to keep the universe ionized. At this time, known as {\it recombination}, the primordial plasma coalesced into neutral atoms, and the mean free path of the photons increased to roughly the size of the observable universe. This radiation has since traveled essentially unhindered through the universe, and provides a snapshot of the universe when it was only $370,000$ years old.
Now, $13.7$ billion years later, the radiation has cooled to microwave frequencies and is observed as the cosmic microwave background (CMB), the thermal afterglow of the Big Bang.

To a very good approximation, the temperature of the CMB is uniform across the whole sky; moreover, it is the most perfect black-body spectrum known, with a mean temperature of $\bar T_0 = 2.725^\circ$~K as measured by the COsmic Background Explorer ({\sl COBE}) satellite \cite{FIRAS} in 1992. The discovery of the CMB \cite{pw65}, together with the black-body nature of its frequency spectrum, was of fundamental importance to cosmology because it validated the idea of a {\it hot Big Bang} -- {\it i.e.}~the universe was hot and dense in the past and has since cooled by expansion\cite{DickeCMB}. Equally important is the fact that the CMB has slight variations of one part in 100,000 in its temperature \cite{DMR}. 
The most accurate 
measurement of these fluctuations is by the Wilkinson Microwave Anisotropy Probe ({\sl WMAP})\cite{Hinshaw:2008kr}.
The {\it temperature anisotropies} reflect the primordial inhomogeneities in the underlying density field that provided the seeds for cosmological structure formation.

But what created these primordial inhomogeneities? In this review we describe how, in the initial moments of the Big Bang, a period of exponential expansion called {\it inflation}\cite{Guth, Linde, Steinhardt} might have caused the universe to expand by at least a factor of $10^{26}$ in an infinitesimal time ($\sim 10^{-33}$ seconds)\cite{Guth, mfb92,star79,star85}. The expansion was driven by a hypothetical quantum field called the {\it inflaton}, which sourced {\it negative pressure} and {\it accelerated expansion}. 
The physical size of the universe grew so much that it became much larger than the distance that light could have traveled since the Big Bang ({\it i.e.}~our observable horizon). Any inhomogeneities that preceded  inflation were erased and the universe became flat and smooth throughout our observable patch, in the same way that the surface of the earth looks flat when viewed from a small aircraft, even though its global shape is spherical. However, the theory also predicts that tiny quantum mechanical fluctuations in the inflaton field resulted in the perturbations imprinted on the CMB and the large scale distribution of galaxies. This is the currently dominant theory for the generation of the initial inhomogeneities.

In this review, we aim to provide a mostly qualitative introduction to the inflationary epoch of the early universe.  We describe the basic predictions of the theory, as well as the latest observational constraints.

\vskip 6pt
The outline of the review is as follows: 
In \S\ref{sec:concordance} we introduce the standard model of cosmology. We describe the homogeneous background dynamics of the universe and the evolution of small fluctuations under the influence of gravity.  We show how, over time, primordial fluctuations grow 
through gravitational instability
to become the large-scale structure of the universe.
In \S\ref{sec:CMB} we explain how fluctuations in the density and the spacetime metric lead to CMB temperature and polarization anisotropies.  We introduce the theory of inflation in \S\ref{sec:inflation}. After explaining how inflation resolves fundamental conceptual problems with the standard Big Bang cosmology, we show that a quantum treatment of the inflationary dynamics provides an elegant mechanism to explain the origin of all structure in the universe.  In \S\ref{sec:observations} we relate the predictions of inflation to current CMB observations.
Finally, in \S\ref{sec:future} we discuss the discovery potential of future CMB experiments.
We conclude in \S\ref{sec:conclusions} that early universe cosmology is headed for an exciting future.

\vskip 6pt
To make this article accessible to a wide audience, many technical details have been suppressed or simplified.
We refer the reader to the large literature on the theory of inflation\cite{LiddleLyth, Mukhanov, WeinbergCosmology, Dodelson} for a more precise treatment.

\vfil

\newpage
\section{Concordance Cosmology}
\label{sec:concordance}

\subsection{The Homogeneous Universe}

There is undeniable evidence for the expansion of the universe:  the light from distant galaxies is systematically shifted towards the red end of the spectrum \cite{Hubble}, the observed abundances of the light elements (H, He, and Li) match the predictions of Big Bang Nucleosynthesis (BBN) \cite{abc}, and the cosmic microwave background can only be explained as a relic radiation from a hot early beginning \cite{DickeCMB}. 

The expansion of the universe is understood in the context of General Relativity, Einstein's theory of gravitation.
It is an observational fact that the universe is homogeneous and isotropic when averaged over cosmological
scales.
The spacetime geometry is then determined by a single function of time, the scale factor $a(t)$, that characterizes the expansion of homogeneous three-dimensional spatial slices
\beq
\label{equ:metric}
\d s^2 \equiv g_{\mu \nu} \d x^\mu \d x^\nu = - \d t^2 + a(t)^2 \d {\bf x}^2\, .
\eeq
  The distance between any two objects in the universe is proportional to $a(t)$. The redshift $z$ (the fractional change in the wavelength of light between the time it is emitted and the time it is observed)
 is proportional to $1/a(t)$  
 -- {\it i.e.}~larger values of $z$ (smaller values of $a(t)$) correspond to larger distances or earlier times; see Fig.~\ref{fig:timeline}.  Einstein's equations determine the time evolution of the scale factor $a(t)$ in terms of the energy content of the universe.
Assuming that the universe is filled with a homogeneous fluid with energy density $\rho$ and pressure $p$, the scale factor satisfies the following equations
\beq
\label{equ:Friedmann}
H^2 \equiv \Bigl( \frac{1}{a} \frac{d a}{d t}\Bigr)^2 = \frac{8\pi G}{3} \rho  - \frac{k}{a^2}\, ,
\eeq
and
\beq
\frac{1}{a} \frac{d^2 a}{d t^2} = - \frac{4\pi G}{3} (\rho + 3 p) \, .
\eeq
A combination of cosmological observations, including measurements of CMB fluctuations\cite{boomerang,Hinshaw:2008kr,Reichardt:2008ay,Readhead:2004gy,Grainge:2002da}, the distribution of galaxies\cite{Tegmark:2003uf,Cole:2005sx,Tegmark:2006az},  and distances to  type Ia supernova explosions\cite{Kowalski:2008ez,Frieman:2008sn}, has revealed a universe filled with photons, baryons (4\%), dark matter (23\%), and dark energy (73\%).  The spatial curvature of the universe $k/a^2$ is found to be negligible.

\subsection{From Primordial Fluctuations\\ to Large-Scale Structure}

The homogeneous universe of the previous section can only be an approximation; our own existence proves that the universe is not perfectly homogeneous (and never has been). Small inhomogeneities have to exist in order to explain the formation of galaxies, planets and ultimately life. These inhomogeneities are parameterized as spatial fluctuations in the density relative to the homogeneous background density $\bar \rho(t)$,
\beq
\delta \rho(t, {\bf x}) \equiv \rho(t, {\bf x}) - \bar \rho(t)\, .
\eeq
The function $\delta \rho(t, {\bf x})$ describes a complicated ``landscape" of density fluctuations with peaks of high density and valleys of low density.  One is typically not interested in the precise shape of this density landscape, but prefers to describe it statistically, {\it e.g.}~one studies the correlations between density fluctuations at two different points in space, $\langle \delta \rho(0) \delta \rho(\Delta {\bf x}) \rangle$. This answers questions such as: what is the probability of finding a high density peak at a distance $\Delta {\bf x}$ from another high density peak? The two-point correlation in real space corresponds to the {\it power spectrum} in Fourier space,\footnote{For a flat universe, Fourier space corresponds to a plane wave decomposition of $\delta \rho(t, {\bf x})$. We denote the comoving wavenumber of each plane wave by $k$.}
\beq
P_s(k) = A_s (k/k_\star)^{n_s -1}\, .
\label{eq:pk_scalar}
\eeq
Here, we have defined a simple parameterization of the power spectrum in terms of an amplitude $A_s$ and a spectral index $n_s$, both measured at an arbitrary reference scale $k_\star$.
Much of observational cosmology is concerned with measuring the power spectrum of density inhomogeneities and its time-evolution. Models of inflation (\S\ref{sec:inflation}) predict the detailed shape of the function $P_s(k)$ which can then be compared with observations (\S\ref{sec:observations}).

\begin{figure}[htbp!]
    \centering
        \includegraphics[width=.45\textwidth]{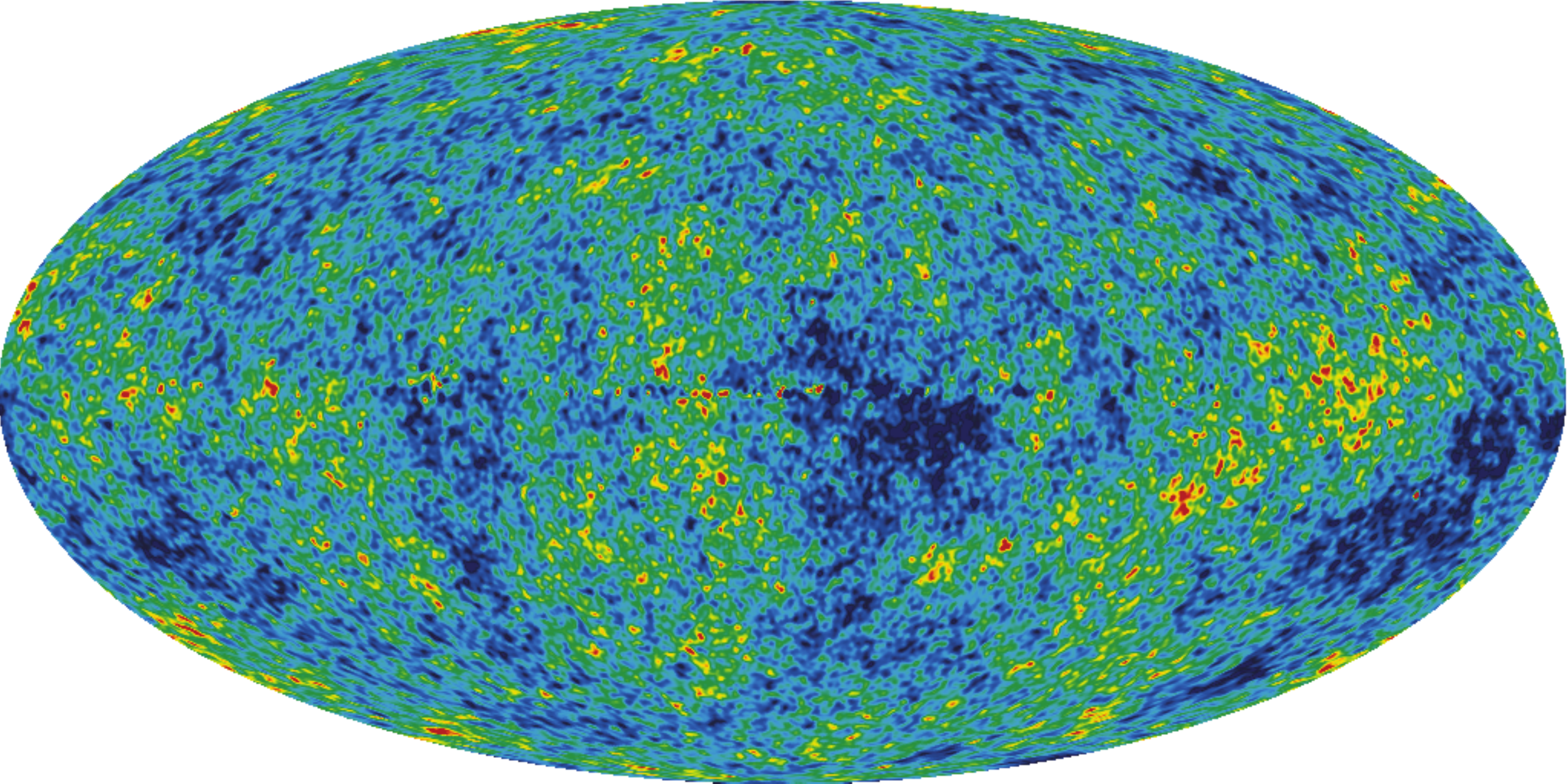}
    \caption{\small Temperature fluctuations in the cosmic microwave background (CMB). Blue spots represent 
 line-of-sight
    directions on the sky where the CMB temperature is $\sim 10^{-5}$ below the mean, $\bar T_0=2.725^\circ$~K.  This corresponds to photons losing energy while climbing out of the gravitational potentials of overdense regions in the early universe.  Yellow and red indicate hot (underdense) regions.  The statistical properties of these fluctuations contain important information about both the background evolution and the initial conditions of the universe (see Figures \ref{fig:PowSpec} and \ref{fig:CMBparam}). This figure is reproduced from Hinshaw et al.~(2009) ApJS, Vol 180, with permission from the AAS.}
    \label{fig:cmb}
\end{figure}

An essential aspect of the theory of cosmological density fluctuations is the fact that  density fluctuations grow over time via gravitational instability. High density regions continuously attract more matter from the surrounding space. This increases the density of the high density regions while depleting the surrounding low density regions. At late times the highest density peaks collapse into the large-scale structure of the universe. Today we observe this effect in the clustering properties of galaxies on the sky. At early times, density fluctuations leave imprints in the temperature of the cosmic microwave background (\S\ref{sec:CMB}) (see Fig.~\ref{fig:cmb}).

We also expect the universe to be filled with a stochastic background of
gravitational waves, {\it i.e.}~fluctuations in the background spacetime (\ref{equ:metric}),
\beq
\label{equ:hij}
h_{ij}(t, {\bf x}) = g_{ij}(t, {\bf x}) - \bar g_{ij}(t)\, .
\eeq
Again, a compact statistical description is the power spectrum,
\beq
P_t(k) = A_t (k/k_\star)^{n_t}\, .
\label{eq:pk_tensor}
\eeq
Gravitational waves interact only very weakly with matter, but they leave subtle imprints in the polarization of the CMB (\S\ref{sec:CMB}).
The quest for a detection of this signature of primordial gravitational waves in the CMB polarization signal is
central to the future of observational cosmology (see \S\ref{sec:future}).

\section{The Cosmic Microwave Background}
\label{sec:CMB}

\subsection{Temperature Anisotropy}

Fig.~\ref{fig:cmb} shows the latest map of the microwave sky in a Mollweide projection\cite{Hinshaw:2008kr}. Fluctuations in the CMB temperature have been color-coded with red (blue) spots corresponding to 
 line-of-sight
directions on the sky where the CMB temperature is $\sim 10^{-5}$ above (below) the mean, $\bar T_0 = 2.725^\circ$~K. Directions of low CMB temperature correspond to photons losing energy while climbing out of the gravitational potentials of overdense regions in the early universe. The physics governing the evolution of the temperature anisotropies in the CMB is very well understood and will be described below. First, however, we will describe the statistical techniques used to analyze the CMB maps.

\vskip 6pt
\noindent
{\it Power spectrum}

\begin{figure}[htbp!]
    \centering
        \includegraphics[width=.45\textwidth]{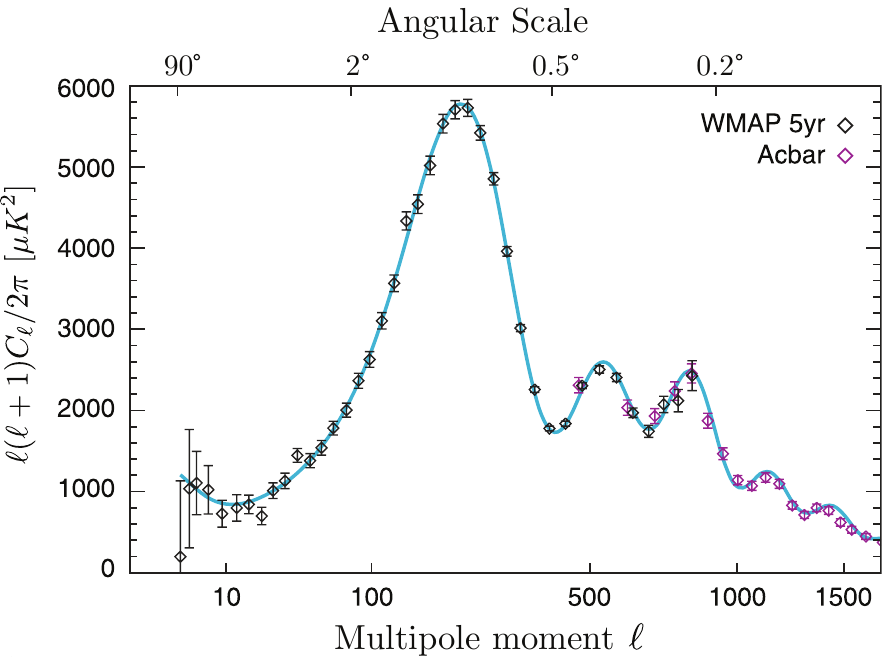}
   \caption{Power spectrum of CMB temperature fluctuations. {\it Blue line:} prediction of the $\Lambda$CDM concordance model with a nearly-scale invariant input spectrum. This figure is reproduced from Hinshaw et al.~(2009) ApJS, Vol 180, with permission from the AAS.}
    \label{fig:PowSpec}
\end{figure}

 We express the CMB temperature fluctuations, $\Delta T(\theta,\phi) \equiv T -\bar T_0$,  in terms of a spherical harmonic expansion
\begin{equation}
\Delta T(\theta,\phi)=\sum_{\ell, m} a_{\ell m} Y_{\ell m}(\theta, \phi)\, ,
\end{equation}
where ($\theta$, $\phi$) denote position
angles on the sky.
If the anisotropies form a Gaussian random field\footnote{Below we discuss the theoretical motivation and the observational evidence for this assumption.}, then all
the statistical information in the CMB map is contained in the {\it
angular power spectrum}\footnote{This is the harmonic transform of the two-point correlation function $\langle \Delta 
T(\hat{n}) \Delta T(\hat{n}') 
\rangle$, where $\hat{n}$ and $\hat{n}'$ are unit vectors representing different lines of sight.}:
\begin{equation}
C_{\ell}=\frac{1}{2\ell+1}\sum_{m}|a_{\ell m}|^2\;.
\end{equation}
This describes the cosmological information contained in the
millions of pixels of a CMB map in terms of a much more compact
data representation. 
The power spectrum of the temperature anisotropies shown in Fig.~\ref{fig:cmb} is given in Fig.~\ref{fig:PowSpec}.  The agreement between the theory and the data is remarkable.

We see that
the power spectrum of the CMB temperature anisotropies contains three distinct regimes --- (1) at
angular separations larger than $\sim 2^\circ$, 
there is a relatively flat plateau; (2) at intermediate angular
scales, between $\sim 2^\circ$ and a few arcminutes, there is a series
of peaks and troughs which are (3) exponentially damped at
sub-arcminute scales. These features can be understood in
terms of a simple physical analogy. 

\vskip 6pt
\noindent
{\it Cosmic sound waves}

Just before recombination, the universe contained a tightly-coupled
photon-baryon fluid, along with dark matter which
was not coupled to the photon-baryon fluid since it does not
participate in electromagnetic interactions. There were tiny
perturbations in the density, and hence gravitational potentials,
over a wide range of length-scales. The perturbations in the dark
matter grow continuously as the universe expands, but the
gravitationally-driven collapse of perturbations in the photon-baryon
fluid is resisted by the pressure of the photons in the fluid, which
acts as a restoring force. This means that an overdensity in the
photon-baryon fluid falls into a potential well, is compressed till
the collapse is halted by the radiation pressure exerted by the photons, and
then rebounds till the expansion is halted by the weight of the fluid
and the gravity of the dark matter potential well, causing the mode to
recollapse once more. In short, the tug of war between gravity and
pressure sets up a sound wave in the photon-baryon fluid, and the
physics is essentially described by a forced harmonic oscillator
\cite{hs95}, with the overall  amplitude of the wave-form being fixed
by the initial amplitude of the perturbations. On sub-degree scales,
smaller than the sound horizon at the surface of last-scattering (SLS), we observe modes
which have had time to undergo this oscillation. Since last-scattering
is nearly instantaneous, modes with different wavelengths are caught
at different phases in their oscillation. Modes that were at maximum
compression or rarefaction correspond to {\it peaks} of the CMB power
spectrum (which is the {\it square} of the amplitude), while the {\it
troughs} correspond to velocity maxima (where the densities correspond
to neutral compression, {\it i.e.}~the velocity is out of phase by $\pi/2$
with respect to the density). In terms of a map of the CMB temperature
anisotropies, on very large scales, a compression (overdensity) makes a cold spot, 
since a photon loses energy and redshifts in climbing out of the extra deep
potential well; and a rarefaction (underdensity) makes a hot spot
compared to the average CMB temperature, since the photon does not
lose as much energy in climbing out of a shallower potential well.

\vskip 6pt
\noindent
{\it Photon diffusion}

This perfect fluid approximation breaks down when photon diffusion
starts to become important. On scales smaller than the photon mean
free path, photons free-stream out of overdensities, erasing the
perturbation. The equation becomes a forced harmonic oscillator with a
friction term representing viscous damping, and the temperature
fluctuations are exponentially damped on sub-arcminute scales
\cite{silk68}.

\vskip 6pt
\noindent
{\it Large-scale plateau}

In addition to these hydrodynamical effects, temperature anisotropy
arises on the very largest scales ($\theta > 2^\circ$), from modes with wavelengths larger
than the sound-crossing distance at last-scattering. These modes are
essentially unaffected by causal physics, frozen in their initial
configurations. In this case, purely general relativistic effects create temperature
anisotropy on these large scales; for example, the so-called
Sachs--Wolfe effect \cite{sw67} changes the CMB
temperature as photons traverse time-varying
gravitational potentials. This happens when gravitational potentials
decay at early times when radiation dominates and at late times when
dark energy dominates.

\vskip 6pt
\noindent
{\it Projection effects}

Of course we do not directly observe the sound waves at the SLS; what we
measure is their projection on the sky. Since the sound crossing scale
at the SLS is a well-defined quantity, a ``standard ruler'', the angle $\theta_s$ that
it subtends is an important probe of  the geometry of the space between the
observer  and the SLS.
If the spatial geometry of the universe were flat
(Euclidean), the photons would follow a simple straight-line path from
the SLS to us. If it were positively curved, such as the surface of
a sphere, the photons follow a curved trajectory (such as the lines of
longitude on the surface of the Earth) and $\theta_s$ is larger than in a
flat universe. If it were negatively curved (saddle-shaped), $\theta_s$ is
smaller. 
Since, according to General Relativity, the
matter-energy content shapes the geometry of spacetime, a
precise measurement of $\theta_s$ tells us about the total matter-energy
density content of the universe.

\begin{figure}[h!]
    \centering
        \includegraphics[width=0.45\textwidth]{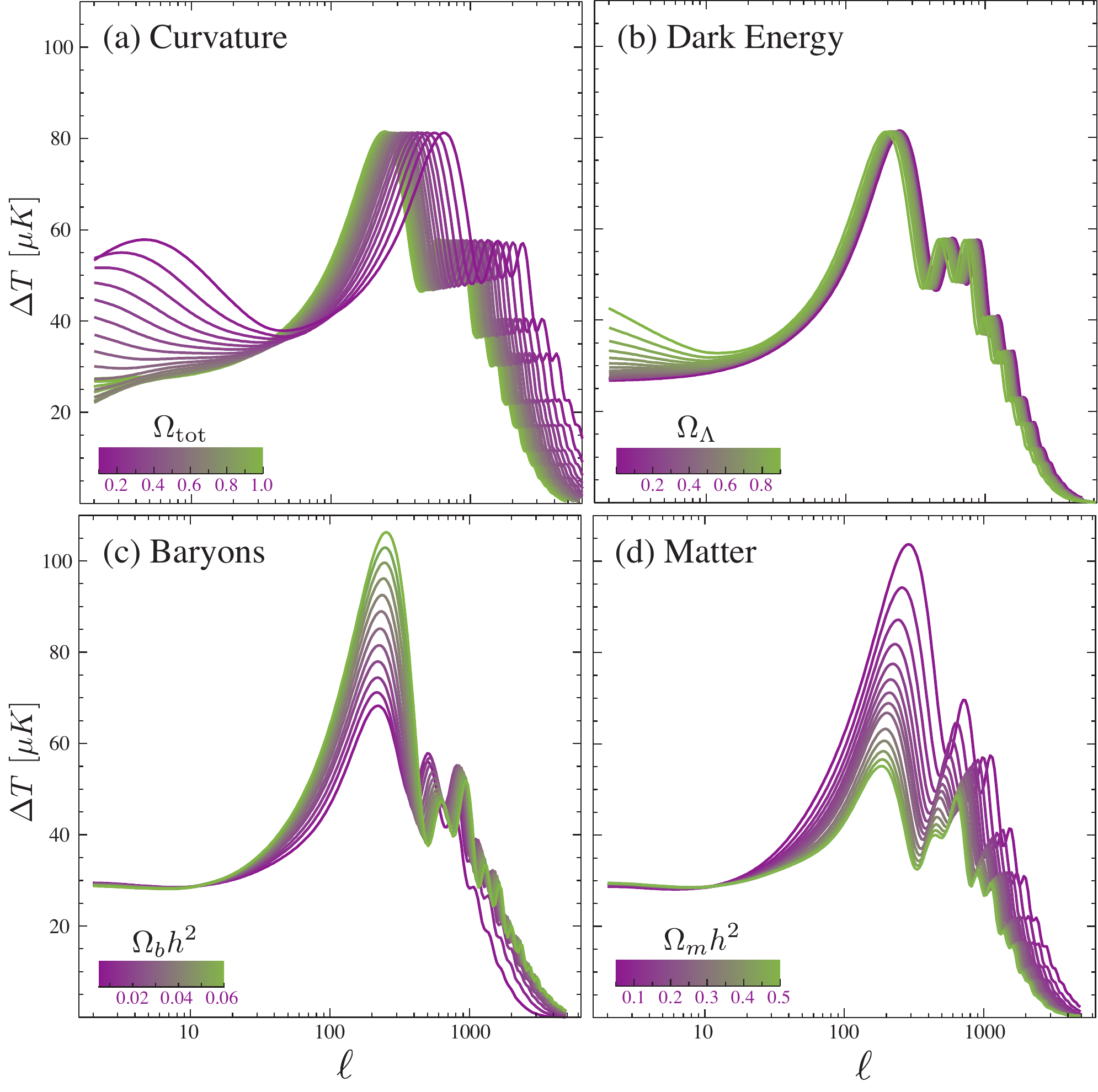}
    \caption{\small Fluctuations in the cosmic microwave background as a function of the parameters of the background cosmology {\scriptsize [figure courtesy of Wayne Hu]}.
    \newline
    (a) Variation of the total density (or curvature) shifts the positions of the peaks of the spectrum.  The CMB is therefore a probe of the background geometry. (b) Increasing the dark energy contribution increases power on large scales via the integrated Sachs-Wolfe effect.  (c) The baryon density affects the relative peak heights. The observed relative peak heights are consistent with Big Bang nucleosynthesis values for the baryon density.  (d) Increasing the matter content (dark matter and baryons) uniformly damps power on all sub-horizon scales. This figure is reprinted with permission from  Hu, W. \& Dodelson, S.~(2002) \copyright\ ARA\&A, Vol 40, p.~137, {\tt http://www.annualreviews.org}.}
    \label{fig:CMBparam}
\end{figure}

\vskip 6pt
\noindent
{\it Cosmological parameters}

We just described how projection effects are sensitive to the global curvature of spacetime.
In Fig.~\ref{fig:CMBparam} we illustrate how further details of the CMB power spectrum (like the peak positions and relative peak heights) allow a measurement of the matter budget of the universe.
These observations (together with other cosmological probes like supernova distances) have revealed a universe filled with atoms (4\%), dark matter (23\%) and dark energy (73\%).

\subsection{Polarization Anisotropy} \label{polani}

\noindent
{\it Polarization from Thomson scattering}

\begin{figure}[htbp!]
    \centering
        \includegraphics[width=0.4\textwidth]{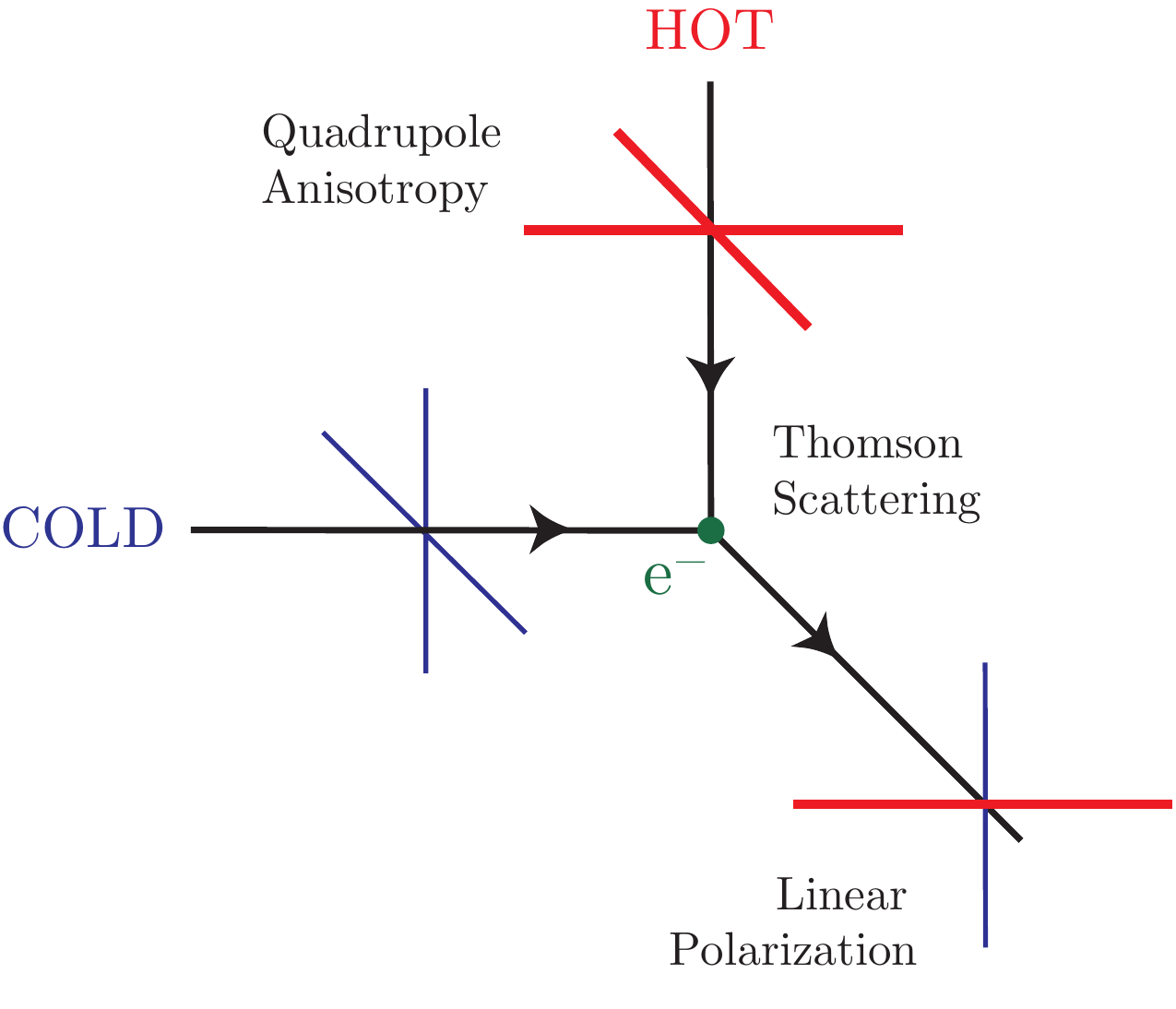}
   \caption{Thomson scattering of radiation with a quadrupole anisotropy generates linear polarization.  Red colors (thick lines) represent hot and blue colors (thin lines) cold radiation \cite{HuWhite}. }
    \label{fig:pol}
\end{figure}

The CMB is not only characterized by temperature fluctuations, but also by polarized anisotropies\cite{rees68,kai83}. The cross-section for Thomson scattering of photons by electrons depends on the polarization states of the incoming and outgoing radiation. If a free electron
``sees'' an incident radiation pattern that is isotropic, then the
outgoing radiation remains unpolarized because orthogonal polarization
states cancel out. However, if the incoming radiation field has a
quadrupolar anisotropy, a net linear polarization is generated (see Fig.~\ref{fig:pol}). 
Since quadrupolar temperature anisotropy is
generated at last-scattering when the tight-coupling approximation breaks down, linear polarization results from the relative velocities of
electrons and photons on scales smaller than the photon diffusion
length-scale. Since both the velocity field and the temperature
anisotropies are created by density fluctuations, a component of the
polarization should be correlated with the temperature anisotropy. An
important corollary to this argument is that no additional
polarization anisotropy is generated after last-scattering, since
there are no free electrons to scatter the CMB photons ({\it i.e.}~there is no equivalent of the Sachs-Wolfe effect for polarization). However, when
the first generation of stars forms, their UV light reionizes
the universe; free electrons scatter CMB photons, introducing some
optical depth and uniformly suppressing the power spectrum of the
temperature fluctuations by $\sim 30\%$. Furthermore, the free electrons see the local
CMB quadrupole at the redshift of star-formation and polarize the CMB
at large scales where no other mechanism of polarization operates
\cite{zal97}. The CMB polarization anisotropy is small -- in the
standard picture of the thermal history, it is only a few percent of
the temperature anisotropy. Consequently it is much harder to measure
than the former.

\vskip 6pt
\noindent
{\it B-modes and gravitational waves}

\begin{figure}[htbp!]
    \centering
        \includegraphics[width=.35\textwidth]{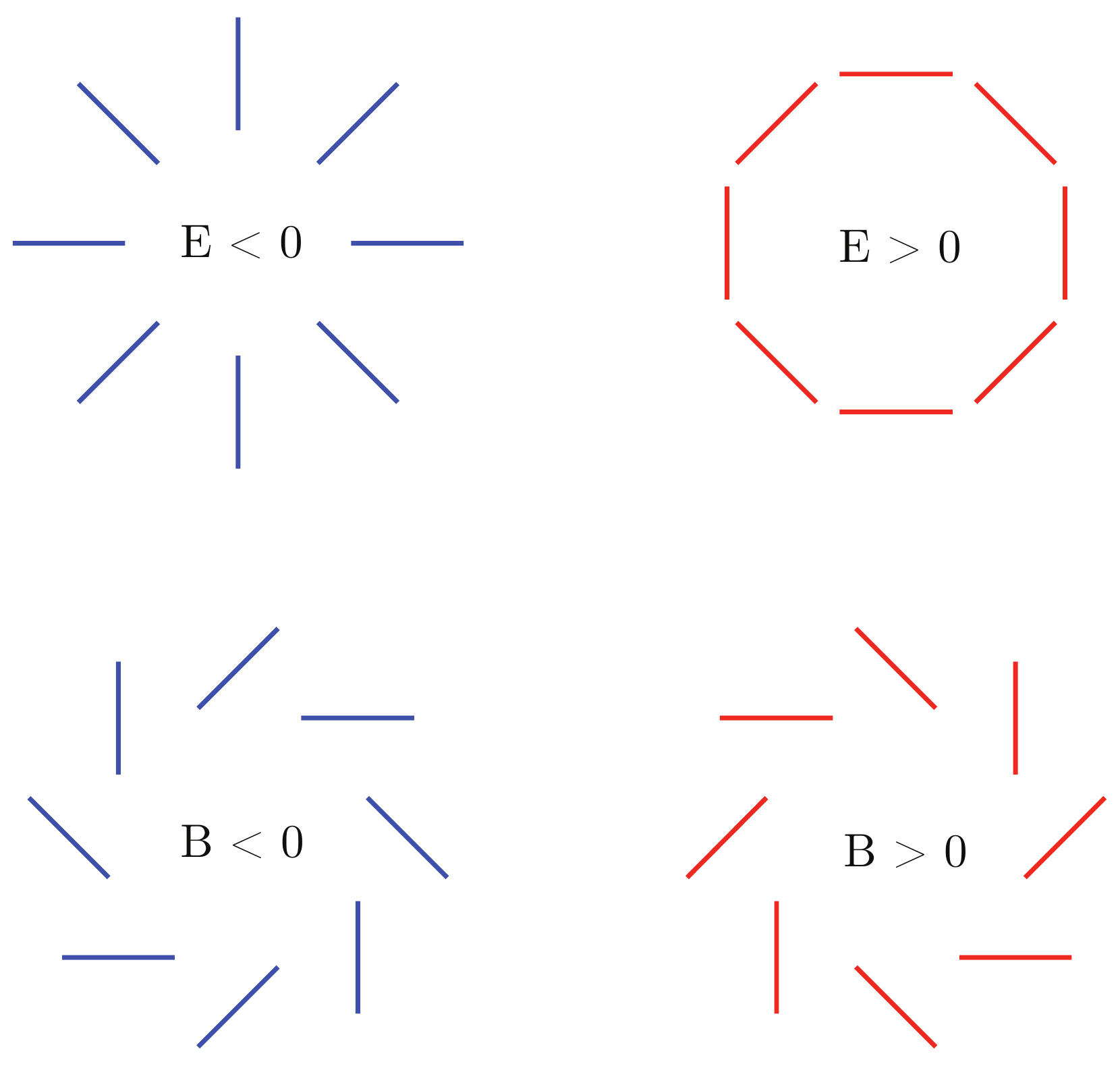}
   \caption{Examples of $E$-mode and $B$-mode patterns of polarization. Note that if reflected across a line going through the center the $E$-patterns are unchanged, while the positive and negative $B$-patterns get interchanged.}
    \label{fig:EBmode}
\end{figure}

For the purpose of this review, a crucial feature of
the CMB polarization anisotropy is its potential to reveal the
signature of primordial gravitational waves \cite{sz97,kks97}. The CMB
temperature anisotropy, being a scalar quantity, cannot
differentiate between contributions from density perturbations (a
scalar quantity) and gravitational waves (a tensor quantity). However,
polarization has ``handedness'', and thus can discriminate between the
two. For this purpose it is useful to decompose the polarization anisotropy into two orthogonal
modes (see Fig.~\ref{fig:EBmode}): 
\begin{enumerate}
\item[i)] {\it E-mode}: a curl-free mode (giving polarization vectors
that are radial around cold spots and tangential around hot spots on the sky) is generated
by both density and gravitational wave perturbations; 
\item[ii)] {\it B-mode}: a
divergence-free mode (giving polarization vectors with vorticity
around any point on the sky) can only be produced by
gravitational waves. 
\end{enumerate}

The primordial $B$-mode anisotropy is at least an
order of magnitude smaller than the $E$-mode polarization. This,
combined with the difficulty of separating primordial $B$-modes from
$B$-modes created by astrophysical foregrounds like polarized dust in our galaxy, makes the measurement of a primordial gravitational wave contribution a great
challenge -- the ``holy grail'' of CMB measurements.

\begin{figure}[htbp!]
    \centering
        \includegraphics[width=.45\textwidth]{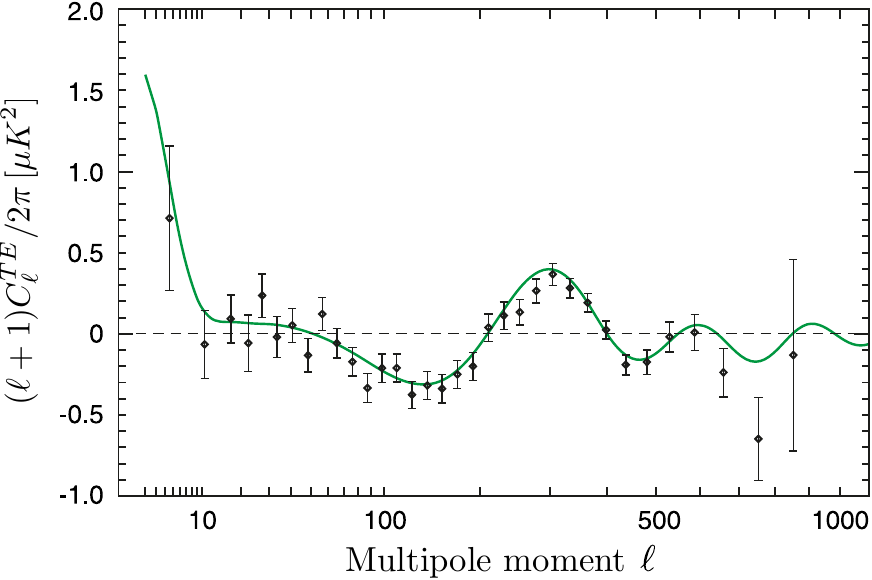}
   \caption{Power spectrum of the cross-correlation between temperature and $E$-mode polarization \cite{WMAP5}. The anti-correlation on scales $\ell = 100-150$ (corresponding to angular separations $\theta > 2^\circ$) is a distinctive signature of adiabatic fluctuations on superhorizon scales at the epoch of decoupling, confirming a fundamental prediction of the inflationary paradigm (see \S\ref{sec:inflation}).  This figure is reproduced from Hinshaw et al.~(2009) ApJS, Vol 180, with permission from the AAS.}
    \label{fig:TE}
\end{figure}

\newpage
\noindent
{\it Power spectra}

The symmetries of temperature and polarization ($E$- and $B$-mode) anisotropies allow four types of correlations:
the autocorrelations of temperature fluctuations and of $E$- and $B$-modes denoted by $TT$, $EE$, and $BB$, respectively, as well as the cross-correlation between temperature fluctuations and $E$-modes: $TE$. All other correlations ($TB$ and $EB$) vanish for symmetry reasons. In Fig.~\ref{fig:PowSpec}  we showed the $TT$ spectrum, while Fig.~\ref{fig:TE} gives the $TE$ cross-correlation.  The $EE$ spectrum has now begun to be measured, but the errors are still large.  So far there are only upper limits on the $BB$ spectrum, but no detection.

The dependence on cosmological parameters of each of these spectra differs, and hence a combined measurement of all of them greatly improves the constraints on cosmological parameters by giving increased statistical power,
removing degeneracies between fitted parameters, and aiding in discriminating between cosmological models.

\section{Inflation} 
\label{sec:inflation}

In the previous section we explained the physics of the cosmic microwave background and described the statistical analysis of maps of the temperature fluctuations.
However, so far we didn't 
consider the origin of the primordial fluctuations that underlie these observations.
In this section, we show how small quantum fluctuations during a phase of accelerated expansion in the very early universe provide an elegant mechanism to dynamically explain the seeds of all structure in the universe.

\vfil

\subsection{Problems of Standard Big Bang Cosmology}

The standard Big Bang cosmology is fantastically successful at explaining the basic characteristics of the observed universe (BBN, CMB, etc.). However, on closer inspection there remain key conceptual puzzles.

\vskip 6pt
\noindent
{\it Flatness problem}

In General Relativity, the geometry of the universe is related to its matter content. If the total energy density of the universe takes the critical value $\rho_c \equiv 3 H^2/(8\pi G)$ ({\it cf.}~Eqn.~(\ref{equ:Friedmann})), then the spatial geometry is flat. For $\rho > \rho_c$ the spacetime is positively curved (like the two-dimensional surface of a sphere), while for $\rho < \rho_c$ it is negatively curved (like the two-dimensional surface of a saddle).
It is conventional to use the parameter $\Omega$ to denote the ratio of the actual energy density of the universe relative to the critical value, {\it i.e.}~$\Omega \equiv \rho/\rho_c$.  The value $\Omega = 1$ therefore corresponds to a flat universe, while $\Omega > 1$ and $\Omega < 1$ denote positive and negative curved spacetime, respectively.
{\it A priori}, General Relativity allows any value for $\Omega$. However,
observations show that the present universe is very nearly flat, $\Omega(t_0) \sim {\cal O}(1)$.

Moreover, in standard Big Bang cosmology, $\Omega = 1$ is an {\it unstable} solution:
any slight difference of $\Omega$ from unity in the early universe will rapidly grow, {\it e.g.}~if $\Omega$ were 0.9 at 1 second after the Big Bang, it would be only $10^{-14}$ today;
if $\Omega$ were 1.1 at 1 second, then it would grow so rapidly that the universe would have recollapsed just 45 seconds later \cite{Guth:2005zr}.
 To explain the geometric flatness of space today therefore requires an extreme fine-tuning in the Big Bang cosmology without inflation.

\vskip 6pt
\noindent
{\it Horizon problem}

Observations of the cosmic microwave background imply the existence  of temperature correlations across distances on the sky that corresponded to super-horizon scales at the time when the CMB radiation was released.  In fact, regions that in the standard Big Bang theory would be causally connected on the surface of last-scattering correspond to only an 
angular separation of order $1^\circ$ on the sky.  However, in Fig.~\ref{fig:cmb} we see that the CMB has nearly the same temperature in all directions on the sky. Yet there was no way to establish thermal equilibrium if these points were never in causal contact before last-scattering.

Two facts are fundamental to an understanding of the horizon problem and its resolution:
\begin{enumerate}
\item[i)] the physical wavelength of fluctuations is stretched by the expansion of the universe,

\item[ii)] the physical horizon ({\it i.e.}~the spacetime region in which one point could affect or have been affected by other points) is time-dependent.
\end{enumerate}
In standard Big Bang cosmology (without inflation), the {\it physical horizon grows faster than the physical  wavelength} of perturbations. This implies that the largest observed scales today were outside the horizon at early times. Quantitatively, according to the standard Big Bang theory, the CMB at the surface of last-scattering should have consisted of about $10^4$ causally disconnected regions. However, the observed near-homogeneity of the CMB tells us that the universe was quasi-homogeneous at the time of last-scattering.  In the standard Big Bang theory, this uniformity of the CMB has no explanation and must be assumed as an
(extremely fine-tuned)
initial condition.

\subsection{Solution of the Big Bang Problems}

Inflationary cosmology is based on the 
hypothesis that the early universe expanded exponentially quickly for a fraction of a second. During inflation the rate of expansion was accelerating and
a small homogeneous patch not bigger than $10^{-26}$ m (orders of magnitudes smaller than an atomic nucleus) grew within about $10^{-35}$ seconds to macroscopic size of order 1 meter.
Eventually the acceleration stopped and the expansion slowed down to the more moderate rate that has characterized our universe ever since.
The 1 meter patch grew to become the observable universe.

We will now demonstrate how this brief period of accelerated expansion solves the problems of the standard Big Bang cosmology:

\vskip 6pt
\noindent
{\it Resolution of the flatness problem}

During a period of accelerated expansion, $\ddot a >0$, a flat universe $\Omega = 1$ becomes an attractor solution. If inflation lasted for at least 60 $e$-folds ({\it i.e.}~the scale factor grew by at least a factor of $e^{60}$ during inflation), then $\Omega$ is driven so close to 1 that we will still observe it near 1 today (even though $\Omega=1$ is unstable).

\vskip 6pt
\noindent
{\it Resolution of the horizon problem}

During inflation the universe expands exponentially and {\it physical wavelengths grow faster than the horizon}.  Fluctuations are hence stretched outside of the horizon during inflation and re-enter the horizon in the late universe.
Scales that are outside of the horizon at CMB decoupling were in fact inside the horizon before inflation.
The region of space corresponding to the observable universe therefore was in thermal equilibrium before inflation and the uniformity of the CMB is given a causal explanation. A brief period of acceleration therefore results in the ability to correlate space over apparently impossible distances.

\vfil
\subsection{Classical Dynamics}

\noindent
{\it Cosmic acceleration}

How could the expansion of the universe have been accelerating?

Paradoxically, the answer lies in our modern understanding of gravity.
In Einstein's general theory of relativity gravity couples both to mass (or energy) and to pressure.
This is in contrast to Newton's theory of gravity where the gravitational field couples only to mass.
Recall Einstein's equation for the acceleration of the scale factor $a(t)$:
\beq
\label{equ:acc}
\frac{1}{a} \frac{d^2 a}{d t^2} = - \frac{4\pi G}{3} (\rho + 3 p) \, .
\eeq
Notice that both energy density $\rho$ and pressure $p$ appear as sources for $\ddot a$.
Ordinary matter has positive energy density, $\rho >0$, and positive (or zero) pressure, $p \ge 0$.
The minus sign of the source term on the r.h.s.~in equation (\ref{equ:acc}) then determines that the expansion of the universe is strictly decelerating. This is consistent with our intuition about gravity: the mutual attraction of all matter in the universe causes the expansion to slow down.

\begin{figure}[htbp]
    \centering
        \includegraphics[width=0.35\textwidth]{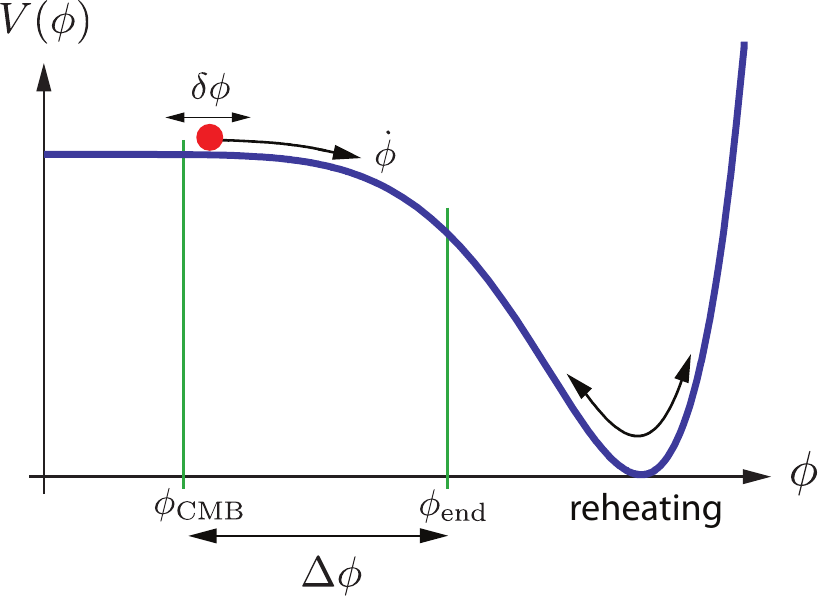}
   \caption{\small Example of an inflaton potential.  Acceleration occurs when the potential energy of the field, $V(\phi)$, dominates over its kinetic energy, $\frac{1}{2} \dot \phi^2$.
Inflation ends at $\phi_{\rm end}$ when the kinetic energy has grown to become comparable to the potential energy, $\frac{1}{2} \dot \phi^2 \approx V$.
CMB fluctuations are created by quantum fluctuations $\delta \phi$ about 60 $e$-folds before the end of inflation. At reheating, the energy density of the inflaton is converted into radiation.}
    \label{fig:small}
\end{figure}

Inflation relies on the early universe being dominated by a very different form of energy.
This is often modeled by a scalar field\footnote{This may be a fundamental scalar field like the Higgs field or it may be a composite field. More generally, we may think of $\phi$ as an order parameter or `clock' measuring the evolution of the energy density during inflation.} $\phi$ (the ``inflaton") with potential energy density $V(\phi)$ (see Fig.~\ref{fig:small}).
One imagines that during inflation the field is displaced from its global vacuum in a state of high energy density called the ``false vacuum".
One further assumes that in the early universe, this scalar field in such a false vacuum state dominates all the contributions to the total energy density.
The energy density of the universe is then determined by the kinetic energy $\frac{1}{2}\dot \phi^2$
and the potential energy $V(\phi)$ of the inflaton field
\beq
\rho = \frac{1}{2} \dot \phi^2 + V(\phi)\, .
\eeq
The pressure associated with the inflaton is the difference of kinetic and potential energy
\beq
p = \frac{1}{2} \dot \phi^2 - V(\phi)\, .
\eeq
If the field evolves 
slowly enough\footnote{For a large value of the potential during inflation, the field is slowed down by Hubble friction: $\ddot \phi + 3 H \dot \phi + V' = 0$,  where $H^2 \sim \frac{8\pi G}{3} V$.}, so that the potential energy dominates over its kinetic energy, $V \gg \frac{1}{2} \dot \phi^2$, then the pressure is {\it negative},
\beq
p \approx - V(\phi) \approx - \rho\, .
\eeq
A slowly evolving scalar field with high but slowly varying potential energy density can therefore source negative pressure.  As we have seen, under these conditions the expansion of the universe 
accelerates and the scale factor grows exponentially:
\beq
a(t) \sim e^{Ht}\, , \qquad H^2 \sim \frac{8\pi G}{3} V\, .
\eeq
 This mechanism to explain acceleration in the early universe is called {\it slow-roll inflation}.

\vskip 6pt
\noindent
{\it The physics of inflation}

\begin{figure}[htbp]
    \centering
        \includegraphics[width=0.35\textwidth]{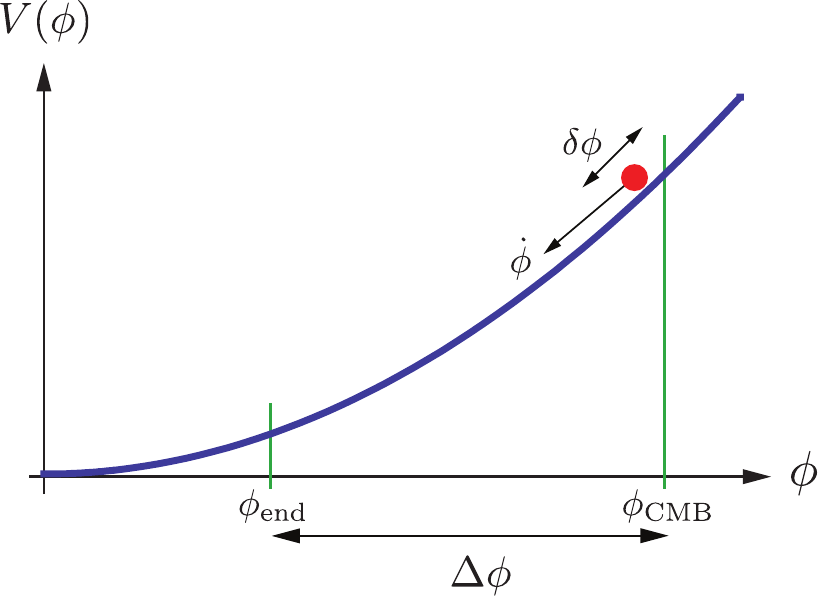}
   \caption{\small Large-field inflation. In an important class of inflationary models the inflationary dynamics is driven by a single monomial term in the potential, $V(\phi) \propto \phi^p$. In these models the inflaton field evolves over a super-Planckian range during inflation, $\Delta \phi > M_{\rm pl}$, and a large amplitude of gravitational waves is produced by quantum mechanical fluctuations.}
    \label{fig:large}
\end{figure}

The fundamental microscopic origin of inflation is still a mystery. Basic questions like: what is the inflaton? what is the shape of the inflaton potential? and why did the universe start in a high energy state? remain unanswered. The challenge to explain the physics of inflation is considerable. Inflation is believed to have occurred at an enormous energy scale $\sim 10^{15}$ GeV, far out of reach of terrestrial particle accelerators. Any description of the inflationary era therefore requires a huge extrapolation of the known laws of physics, and until recently, only a phenomenological parameterization of the inflationary dynamics was possible. In this approach, a suitable inflationary potential function $V(\phi)$ is postulated (see Figures~\ref{fig:small} and \ref{fig:large} for two popular examples) and the experimental predictions are computed from that. As we will see in the next section, details of the primordial fluctuation spectra will depend on the precise shape of the inflaton potential.

\vskip 6pt
\noindent
{\it Inflation in string theory}

Recently, it has become possible to go beyond a simple phenomenological parameterization of the inflaton potential in the context of string theory. In string theory the inflaton field is often given a geometrical interpretation (as the position or orientation of branes or the size or shape of extra dimensions) and the inflaton potential becomes in principle fully computable. Rather than postulating an arbitrary shape for the inflaton potential, in string inflation the inflaton potential is a {\it derived} quantity. 

We refer the reader to the 
comprehensive review by McAllister and Silverstein\cite{EvaLiam} for details on string inflation.
Here, we would only like to comment on the
common misconception that ``anything goes" in the vast ``landscape" of string theory vacua and that the possibilities for string inflation are equally unconstrained.
At present the situation seems quite different.
Microphysical principles (like the need to satisfy Gauss' 
Law on a compact extra-dimensional space) severely restrict the possibilities in inflationary model-building in string theory.
Although unproven, it seems quite likely that string theory models of inflation are more strongly constrained than their quantum field theory counterparts.
One may therefore hope that only a small subset of four-dimensional inflationary actions can arise from consistent string theory constructions.  This would increase the predictive power of inflation.

String inflation is admittedly still in its infancy, but we believe that it holds great promise for revealing the fundamental microscopic origin for the acceleration of the early universe.

\vfil
\subsection{The Quantum Origin of Structure}
\label{sec:origin}

In the previous section we discussed the {\it classical} evolution of the inflaton field. Something remarkable happens when one considers {\it quantum} fluctuations of the inflaton: inflation combined with quantum mechanics provides an elegant mechanism for generating the initial seeds of all structure in the universe. In other words, quantum fluctuations during inflation are the source of the primordial power spectra $P_s(k)$ and $P_t(k)$.
In this section we sketch the mechanism by which inflation relates microscopic physics to macroscopic observables. To simplify the appearance of equations we will work in units where $M_{\rm pl}^{-2} \equiv 8\pi G \equiv 1$.

\vskip 6pt
\noindent
{\it From $\delta \phi$ to $\delta \rho$}

To analyze fluctuations during inflation, the inflaton field may be split into a homogeneous background $\bar \phi(t)$ and a spatially varying perturbation $\delta \phi(t, {\bf x})$
\beq
\phi(t, {\bf x}) = \bar \phi(t) + \delta \phi(t,{\bf x})\, .
\eeq
During inflation the spacetime is quasi-de Sitter space\footnote{de Sitter space is a universe with positive cosmological constant given by the nearly constant inflaton potential $V(\phi)$.}. Perturbations of the inflaton field value $\delta \phi$ satisfy the equation of motion of a harmonic oscillator with time-dependent mass.
The quantum treatment of inflaton perturbations therefore parallels the quantum treatment of a collection of one-dimensional harmonic oscillators.
Just as zero-point fluctuations of a harmonic oscillator induce a non-zero variance for the oscillation amplitude $\langle x^2 \rangle$, the quantum fluctuations during inflation induce a non-zero variance for the inflaton perturbations\cite{Mukhanov} (here given in Fourier space),
\beq
\label{equ:zero}
\langle (\delta \phi_k)^2 \rangle = \frac{V}{6 k^3} \, .
\eeq 
Fluctuations in the inflaton modulate the time when inflation ends (see Figures \ref{fig:small} and \ref{fig:large}). Since the inflaton takes slightly different values at different points in space, inflation ends at slightly different times in different regions of space. This process converts fluctuations in the inflaton $\delta \phi$ into fluctuations in the energy density $\delta \rho$ after inflation. The efficiency of this mechanism for producing cosmological density perturbations depends on the slope of the inflaton potential\footnote{For small $V'$ a given fluctuation $\delta \phi$ leads to a bigger time delay and a larger density fluctuation $\delta \rho$.} $V'$ , while the size of the quantum fluctuations in $\phi$ depend on the scale of the potential $V$.

The full calculation gives the following result for the primordial power spectrum of density fluctuations\cite{LiddleLyth}
\beq
\label{equ:SRPs}
P_s(k) = \left. \frac{1}{12 \pi^2} \frac{V^3}{(V')^2} \right|_{k=a H} \, .
\eeq

\vskip 6pt
\noindent
{\it Gravitational waves}

The two polarization modes of metric perturbations $h_{ij}$ ({\it cf.}~Eqn.~(\ref{equ:hij})) satisfy the same equation of motion as the inflaton perturbation.
Quantum fluctuations in the metric therefore have the variance (\ref{equ:zero}).
The power spectrum of inflationary gravitational waves therefore is\cite{LiddleLyth}
\beq
\label{equ:SRPt}
P_t(k) = \left. \frac{2}{3\pi^2}  V \right|_{k=aH}\, .
\eeq
Observations of $B$-modes of CMB polarization are sensitive to the ratio of tensor to scalar power
\beq
\label{equ:r}
r \equiv \frac{P_t}{P_s} = 8 \left( \frac{V'}{V} \right)^2 \, .
\eeq
As we will explain below, the value of $r$ is of fundamental importance in the quest to understanding the  microscopic origin of the inflationary dynamics.

The power spectra in (\ref{equ:SRPs}) and (\ref{equ:SRPt}) are to be evaluated when a fluctuation with (physical) wavenumber $k/a$ exits the horizon $H^{-1}$.  Different scales exit the horizon at different times when the inflationary potential $V(\phi)$ has slightly different values.  This leads to a small (but computable) scale-dependence of the primordial power spectra. In terms of the parameterization given in equations (\ref{eq:pk_scalar}) and (\ref{eq:pk_tensor}) one finds
\beq
n_s - 1 = 2  \frac{V''}{V}  - 3 \left(\frac{V'}{V}\right)^2\, ,
\eeq
and
\beq
n_t = - 4 \left( \frac{V'}{V}\right)^2\, .
\eeq
Measurements of the scale-dependence of the primordial power spectra therefore probe the shape of the inflaton potential $V(\phi)$.

\subsection{Predictions for the CMB}
 
As we have emphasized throughout this review, the primordial fluctuations predicted by inflation
relate to detailed predictions about the statistics of the hot and cold spots on the CMB sky: how many patches there should be of each angular size, and how much hotter or colder than the average they should be, and so forth. The simplest inflationary scenarios predict the following characteristics for the form of CMB perturbations: 

(1) {\it Flat geometry}, {\it i.e.}~the observable universe should have no spatial curvature. This is measured by the position of the first peak of the CMB power spectrum (Fig.~\ref{fig:cmb}),
combined with a measurement of the present expansion rate (Hubble constant).

(2) {\it Gaussianity}, {\it i.e.}~the primordial perturbations should correspond to Gaussian random variables to a very high precision. 

(3) {\it Scale-invariance}, {\it i.e.}~to a first approximation, there should be equal power at all length-scales in the perturbation spectrum, without being skewed towards high or low wavenumbers. 
In terms of the parameterization (\ref{eq:pk_scalar}) and (\ref{eq:pk_tensor}) this corresponds to $n_s =1$ and $n_t= 0$. However, small deviations from scale-invariance are also a typical signature of inflationary models.

(4) {\it Adiabaticity}, {\it i.e.}~the temperature and matter density perturbations should satisfy the condition $\delta T/ T = \frac13 \delta \rho/\rho$. This is the same correlation found in the adiabatic compression of a gas -- regions of high density are also hotter. It implies that a positive fluctuation in the number density of one species is a positive fluctuation in all the other species. 

(5) {\it Super-horizon fluctuations}, {\it i.e.}~there exist correlations between anisotropies on scales larger than the causal horizon, beyond which two points could not have exchanged information at light-speed during
the history of the universe. This corresponds to angular separations on the sky larger than $\sim 2^\circ$. 
In addition, particular models of inflation predict significant amounts of, 

(6) {\it primordial gravitational waves}, which gives rise to temperature and polarization anisotropies as described above.
In the next section we describe the physical relevance of this final prediction of inflation in more detail.

\subsection{The CMB as a High Energy Experiment}

Given the enormous energy scale at which we believe inflation occurred, 
the physics of inflation is likely to be far out of reach of terrestrial experimentation.
While this makes it particularly challenging to understand the origin of inflation from a particle physics point of view, it conversely may be viewed as a great opportunity to learn about ultra-high energy physics from cosmological observations.

In this section we illustrate the great discovery potential of CMB observations with the example of $B$-mode polarization. 
Two important pieces of information about inflation follow from a $B$-mode detection: 1) the energy scale of inflation, and 2) the field variation of the inflaton.

\vskip 6pt
\noindent
{\it Energy scale of inflation}

From the results of \S\ref{sec:origin} one may derive the following relation between the energy scale of inflation $V^{1/4}$ and the tensor-to-scalar ratio on CMB scales $r$, \beq V^{1/4} = 1.06 \times 10^{16} \, {\rm GeV}
\left(\frac{r}{0.01}\right)^{1/4}\, . \eeq A detectably large tensor amplitude ($r \gtrsim 0.01$) would therefore convincingly
demonstrate that inflation occurred at a tremendously high energy scale, comparable to that of Grand Unified
Theories (GUTs).  It is difficult to overstate the impact of such a result for the high-energy physics
community, which to date only has two indirect clues about physics at this scale: the apparent unification of
gauge couplings, and experimental lower bounds on the proton lifetime.  Some of the earliest successful
inflation models involved direct connections between the inflaton and GUT scale particle physics.  While more
recent models of inflation are usually less tied to our models of particle interactions, instead invoking a
largely modular ``inflation sector", an observed connection between the scale of inflation and the scale of
coupling-constant unification might prompt theorists to re-visit  a possible deeper connection.

\vskip 6pt
\noindent
{\it Super-Planckian field excursion}

From manipulations of equation (\ref{equ:r}) one may derive \cite{Lyth:1996im} the following relation between the tensor-to-scalar ratio $r$ and the distance in inflaton field space between the end of inflation and the point when the scales of CMB fluctuations were created
\beq
\frac{\Delta \phi}{M_{\rm pl}} \gtrsim \Bigl( \frac{r}{0.01} \Bigr)^{1/2}\, .
\eeq
Here, $M_{\rm pl} \equiv (8\pi G)^{-1/2}$ is the reduced Planck mass.
For a definition of $\Delta \phi$ see Figures~\ref{fig:small} and \ref{fig:large}.
A large tensor amplitude, $r>0.01$, therefore correlates with a field variation during inflation that was bigger than Planckian.

Super-Planckian field excursions have interesting theoretical implications\cite{WhitePaper}:
to control the shape of the inflaton potential over a super-Planckian range requires the existence of an approximate shift symmetry in the ultraviolet (UV) limit of the underlying particle theory for the inflaton, $\phi \to \phi + {\rm const}$.
Without a shift symmetry in the UV, large-field inflation is sensitive to an infinite series of Planck-suppressed corrections to the potential, {\it i.e.}~corrections of the form $(\phi/\Mp)^n$.
$B$-modes therefore probe this UV sensitivity of inflation.
In string theory it has only recently become possible to construct controlled large-field inflation models with approximate shift symmetries in the UV\cite{EvaAlex, EvaLiamAlex}.

\section{Current Observational Constraints}
\label{sec:observations}

\subsection{Connecting Primordial Input with Data}

\begin{quote}
{\small 
``It doesn't matter how beautiful your theory is, it doesn't matter how smart you are or what your name is. If it doesn't agree with experiment, it is wrong."  \\
{\it Richard Feynman}}
\end{quote}

\vskip 6pt
\noindent
{\it From $P(k)$ to $C_\ell$}

The CMB is generally regarded as the cosmological data set for which it is easiest to characterize/remove experimental systematics.  In addition, CMB measurements are particularly straightforward to interpret, because the CMB fluctuations are produced by relatively simple, linear physical processes and observed with robust technologies \cite{kss94}. With this knowledge of the underlying physics in hand, one can make precise predictions for the observed temperature anisotropies for a given set of initial conditions and background cosmological model, and compare them to measurements of the CMB temperature and polarization power spectra. In this way, one can deduce the constituents and the initial conditions of the universe, reconstruct its history and evolution, and potentially obtain a window into the epoch when all the structure we see today originated. 

We can illustrate this idea schematically as follows:
\begin{equation}
P_s(k), P_t(k) \to \Delta(T, Q, U) \to C_\ell (TT, TE, EE, BB) \nonumber.
\end{equation}
A given inflationary model predicts the scalar and tensor power spectra $P_s(k)$ and $P_t(k)$ as functions of the wavenumber $k$ (see \S\ref{sec:origin}) which, in combination with a set of cosmological parameters describing the ``late-time''  universe, yields the statistical ensemble of CMB anisotropies $\Delta$ in terms of the temperature intensity $T$ and a polarization vector\footnote{Here, polarization is defined in terms of the Stokes parameters $Q$ and $U$. These are related to $E$- and $B$-modes by a linear transformation.} $(Q,U)$.  If $\Delta$ is a Gaussian random field, then it is fully characterized by the temperature and polarization angular spectra $C_\ell (TT, TE, EE, BB)$. As mentioned above, one of the generic predictions of inflationary models is that the primordial fluctuations  indeed have Gaussian random phases.\footnote{This is a slight oversimplification -- in \S\ref{sec:future} we discuss why small amounts of non-Gaussianity can be an important probe of inflationary physics.} 
 Since the physics that governs the evolution of the temperature and metric fluctuations is linear, the observed temperature fluctuations are also Gaussian.  If we ignore the effects of non-linear physics and astrophysical foregrounds
at late times, then all of the cosmological information in the microwave sky is encoded in the temperature and polarization power spectra.

\vskip 6pt
\noindent
{\it Foregrounds}

There are several expected sources of non-cosmological signals and of non-Gaussianity in the microwave sky, the most significant being Galactic foreground emission, 
radio sources, and galaxy clusters. 
However, these foregrounds can be accounted for by excising the most contaminated regions using sky cuts (typically 15\%--20\% of the full sky) and by subtracting the remaining contamination using multi-frequency observations. The 
multi-frequency approach relies on the fact that foregrounds do not follow a black-body spectrum. In the case of temperature measurements, the process of foreground removal is relatively straightforward because the primordial signal dominates the foreground signals over a relatively wide frequency range. For polarization measurements, the removal of foregrounds is much more challenging, as Galactic foregrounds in fact dominate the primordial signal over all frequencies. Therefore one must make use of detailed multi-frequency templates to clean the foregrounds on large scales, and observe in known ``clean patches'' where Galactic foregrounds are subdominant to access the signal on small scales. We ignore these effects in the following qualitative discussion. In ``real life", a careful characterization of foregrounds is, of course, essential.

\vskip 6pt
\noindent
{\it Statistical analysis}

Cosmological constraints on a given model are primarily {\it inferential}, obtained under the assumption that our observed universe is a realization from an underlying statistical ensemble. Bayesian statistical inference\citep{Jaynes2003} very naturally lends itself to the task of parameter estimation in cosmology.
We let ${C}_{\ell}$ be the power spectrum of the ensemble and 
${C}_{\ell}^{\rm sky}$ the realization on our sky.
 The ultimate goal of the Bayesian analysis is to find a set of parameters that give the best estimate of the ensemble average $\langle {C}_{\ell}\rangle$.
The {\it likelihood function}, ${\cal L}\left(\widehat{C}_{\ell}|{C}_{\ell}^{\rm th}(\vec{\alpha})\right)$, yields the probability of the data given a model and its parameters $\{\vec{\alpha}\}$. Here, $\widehat{C}_{\ell}$ denotes the data -- our best estimator for ${C}_{\ell}^{\rm sky}$ \cite{Hinshaw:2003fc} -- and ${C}_{\ell}^{\rm th}$ is the theoretical prediction for the angular power spectrum.

If foreground removal did not require a sky cut and the instrumental noise were uniform and purely diagonal, then the likelihood function for a CMB experiment (measuring just temperature, for the purposes of illustration) would have the form \citep{bond/jaffe/knox:2000}
\begin{equation} 
-2\ln{\cal L}=\sum_{\ell}(2\ell+1)\left[\ln\left(\frac{{C}_{\ell}^{\rm th}+{N}_{\ell}}{\widehat{C}_{\ell}}\right)+\frac{\widehat{C}_{\ell}}{ {C}_{\ell}^{\rm th} +{N}_{\ell}}-1\right],
\label{eq:exact_like}
\end{equation}
where the effective bias ${N}_{\ell}$ describes the instrumental noise and the resolution of the experiment. Note that ${N}_{\ell}$ and ${C}_{\ell}^{\rm th}$ appear together in equation (\ref{eq:exact_like}) because the noise and cosmological fluctuations have the same statistical properties; they both are Gaussian random fields. Obviously, a much more complex likelihood function is used to model a realistic CMB experiment, but (\ref{eq:exact_like}) captures many of the underlying concepts. For example, at low $\ell$ (large angular scales), the logarithmic term is important, and the ${C}_\ell$ of the realization that describes our universe is more likely to scatter ``low'' than scatter ``high'' compared to the underlying cosmology, whereas at high $\ell$ (small angular scales) the likelihood closely approximates a Gaussian distribution. Even an ideal, noiseless, full-sky experiment is subject to {\it cosmic variance} -- the fundamental uncertainty arising from the fact that for each multipole $\ell$, we can only measure $2\ell +1$ modes on the sky -- leading to an inevitable error $\delta C_\ell/C_\ell = \sqrt{2/(2\ell + 1)}$ in measuring $C_\ell$. Thus, the goal of any CMB experiment is to be cosmic variance-limited over as wide a multipole range as possible.

Following Bayes' Theorem, the {\it posterior probability} of a model given the data is
\begin{equation}
{\cal P}( {\vec{\alpha}}|\widehat{C}_{\ell}) \propto
{\cal L} \left(\widehat{C}_{\ell}|{C}^{\rm th}_{\ell}({ \vec{\alpha}})\right)
{\cal P}\left(\vec{{\alpha}}\right),
\end{equation}
where ${\cal P}(\vec{\alpha})$ describes the {\it prior probability} of the cosmological parameters and we have neglected a normalization factor that does not  affect parameter estimation. Once the choice of priors is specified, our estimator of $\langle {C}_{\ell}\rangle$ is given by ${C}_{\ell}^{\rm th}$ evaluated at the maximum of ${\cal P}(\vec{\alpha}|\widehat{C}_{\ell})$.  The {\it Markov Chain Monte Carlo} (MCMC) technique has become the standard tool in the field to simulate posterior distributions\citep{Christensen:2000ji,Christensen:2001gj,Knox:2001fz,Kosowsky:2002zt,Verde:2003ey}. The MCMC generates random draws ({\it i.e.}~simulations) from the posterior distribution that are a ``fair'' sample of the likelihood surface. From this sample, we can estimate all of the quantities of interest about the posterior distribution (mean, variance, confidence levels). 

\vskip 6pt
\noindent
{\it Primordial input}

Having given a flavor of what is involved in extracting cosmological parameters, we turn now to the parameterization $\{\vec{\alpha}\}$ which includes the primordial input ($P_s$, $P_t$) as well as the late-time cosmological parameters ({\it e.g.} $\Omega_b$, $\Omega_m$, $\Omega_\Lambda$, $H_0$, $\tau$). There are at least three ways to parametrize the primordial sector: i) A standard practice is to assume that the primordial power spectra, which are weakly scale-dependent functions in the simplest inflationary models, are described by an empirical parameterization with an amplitude and a tilt ({\it cf.}~Eqns.~(\ref{eq:pk_scalar}) and (\ref{eq:pk_tensor})). ii) The second approach is to  numerically compute the exact $P_s(k)$ predicted by a given inflationary model\cite{Grivell:1996sr,peirisetal03,Covi:2006ci,Martin:2006rs,Lesgourgues:2007aa,Ringeval:2007am,Lorenz:2007ze,Hall:2007qw,Bean:2007eh,Hamann:2008pb,Ballesteros:2007}, compare this with the data to directly constrain the parameters of that potential, and repeat this for each plausible model, followed by a model-comparison statistic\cite{Bridges:2006,Parkinson:2006,Liddle:2006, Gordon:2007,Trotta:2008,Feroz:2008} to compare the goodness-of-fit to the data. This is a very time-consuming procedure since the number of proposed inflationary models is very large. iii) The third method is the idea of ``reconstruction'': attempting to directly constrain $V(\phi)$ under the assumption that the inflationary paradigm describes the universe\cite{Turner:1993su,Copeland:1993jj,Copeland:1993ie,Liddle:1994cr,Leach:2002ar,Leach:2002dw,Leach:2003us,Kinney:2002qn,Easther:2002rw,Kinney:2006qm,Peiris:2006sj,Peiris:2006ug,Adshead:2008vn}. This reconstruction can just apply to the few $e$-foldings of inflation that describe cosmological scales\cite{Lesgourgues:2007aa,Hamann:2008pb},
or else extrapolate outside this range making use of various physical consistency conditions and observational restrictions\cite{flowroll2008}.

For future, ultra-high-precision cosmological data, using the most optimal statistical techniques in this regard will be extremely important. However, given the quality of the present data, these subtleties of  parameterization are not important, as all these approaches paint the same ``big picture''.

\subsection{History of CMB Observations}

CMB observations have a long history beginning with the accidental discovery of the CMB by Penzias and Wilson\cite{pw65} and culminating in the recent high-precision measurements by the {\sl WMAP} satellite\cite{Hinshaw:2008kr}.
 
In 1965, Arno Penzias and Robert Wilson of Bell Labs worked on a radio telescope and discovered an excess 3.5 $\pm 1^\circ$~K 
 antenna temperature that they couldn't 
 explain\cite{pw65}.  At the time, Penzias and Wilson were completely unaware of the cosmological significance of their measurement.
Dicke, Peebles, Roll, and Wilkinson gave the cosmological interpretation of their signal as the thermal afterglow of the Big Bang\cite{DickeCMB}.
  In 1992, the {\sl COBE} satellite provided a more detailed view of the microwave sky.
{\sl COBE}'s {\sl FIRAS} instrument measured the black-body nature of the CMB spectrum to extreme accuracy\cite{FIRAS}, while its {\sl DMR} instrument provided the first detection of temperature fluctuations on large scales\cite{DMR} ($\theta >7^\circ$).
 A series of ground-based experiments then extended measurements of the spectrum of CMB fluctuations to smaller scales.
In 1998, the balloon experiment {\sl BOOMERanG} detected the first acoustic peak confirming a central prediction of the physics of CMB anisotropies\cite{boomerang} (see \S\ref{sec:CMB}).
The measured position of the first peak indicated that the geometry of the
 universe is flat.
 In 2002, the
 {\sl DASI} experiment announced the first measurement of CMB polarization\cite{DASI}.
 
In recent years, our picture of cosmology, including the early universe, has come into sharp focus mainly due to the {\sl Wilkinson Microwave Anisotropy Probe} ({\sl WMAP}) experiment, which has observed the temperature and polarization anisotropy of the CMB for over five years. {\sl WMAP} was launched on June 30, 2001 from Cape Canaveral and observes the CMB sky from an orbit about the second Lagrange point of the Earth-Sun system, $L_2$.

\begin{figure}[htbp]
    \centering
        \includegraphics[width=0.4\textwidth]{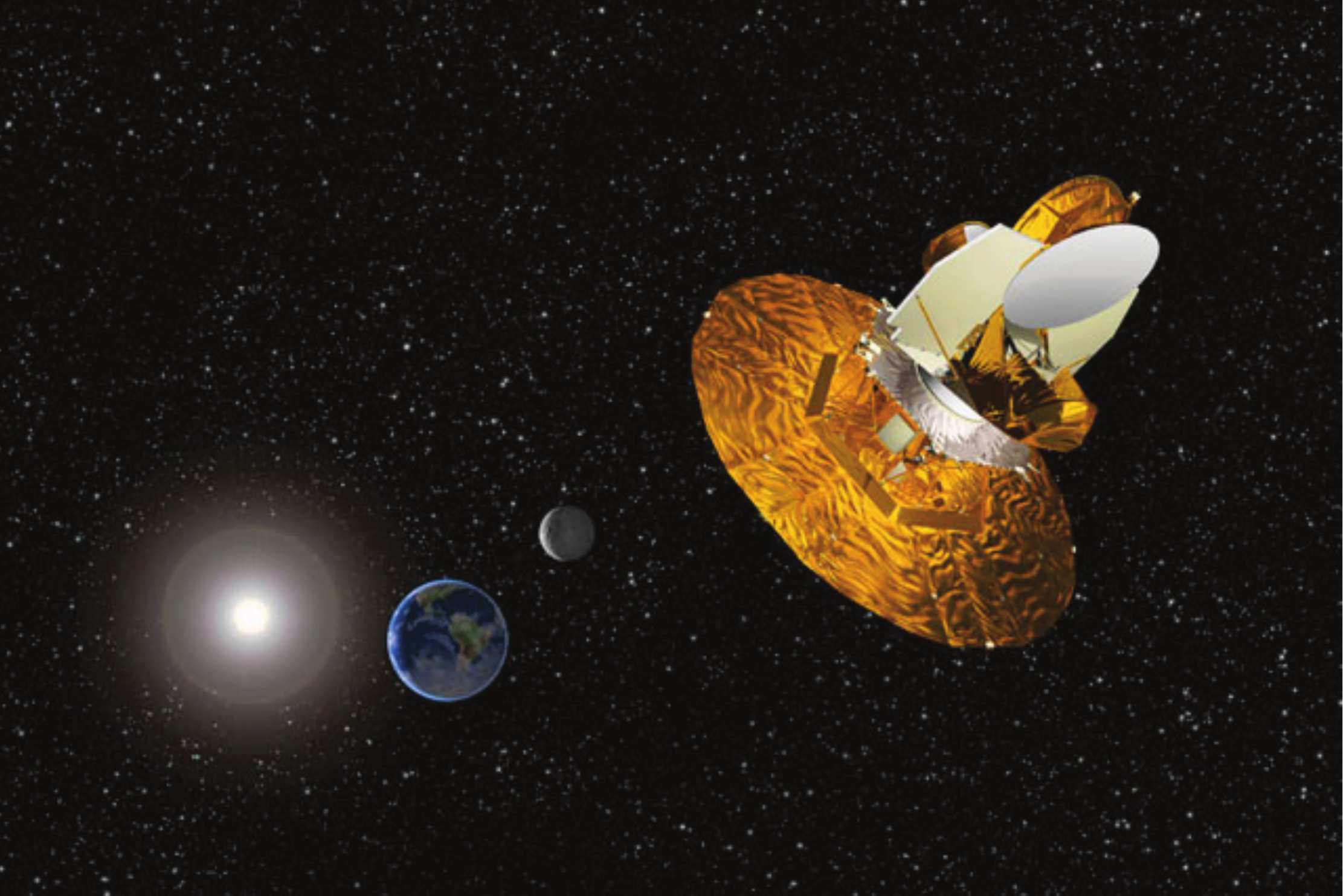}
   \caption{\small 
   Artist's impression of the {\sl WMAP} satellite at $L_2$. Image credit: NASA/WMAP Science Team.}
    \label{fig:wmapS}
\end{figure}

The central design philosophy of the {\sl WMAP} mission was to minimize sources of systematic measurement errors. To achieve this goal, {\sl WMAP} utilizes a differential design; it observes the temperature differences between two directions in the sky, using a back-to-back set of nearly identical optics \cite{pageetal03a}. These optics focus the radiation into horns that feed differential microwave radiometers \cite{jarosiketal03}. Calibration errors are $< 0.5\%$, and the low level of systematic error is extensively 
characterized\cite{pageetal03a, jarosiketal03, hinshawetal03a, barnesetal03,Jarosik:2006ib,Hill:2008hx}. Full-sky maps in five frequency bands from 23--94 GHz are produced from the radiometer data of temperature differences. The resolution of the {\sl WMAP} satellite
is 30 times greater than the previous full sky map by the {\sl COBE} satellite \cite{cobe}. The multi-frequency data enables the separation of the CMB signal from the foreground dust, synchrotron and free-free emission \cite{bennettetal03b, Kogut:2007tq,Gold:2008kp}, since the foreground emission has distinctive frequency dependence while the CMB behaves as a black-body.

\begin{figure*}[t!]
    \centering
        \includegraphics[width=.8\textwidth]{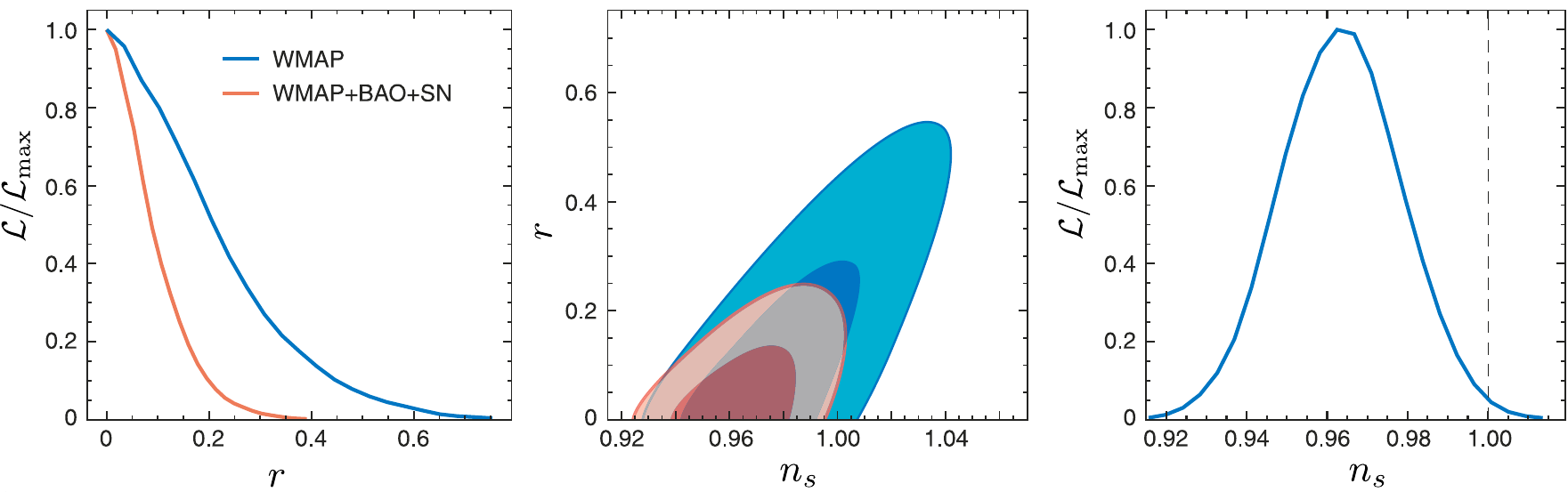}
   \caption{\small {\sl WMAP} 5-year constraints on the inflationary parameters\cite{WMAP5} $n_s$
 and $r$. The {\sl WMAP}-only results are shown in blue, while
 {\sl WMAP} and other observables (Baryon Accoustic Oscillations (BAO) and Supernovae (SN)) are in red. This figure is reproduced from Komatsu et al.~(2009) ApJS, Vol 180, with permission from the AAS.}
    \label{fig:WMAP5}
\end{figure*}

\subsection{{\sl WMAP} and Inflationary Physics}

The {\sl WMAP} satellite has measured the $TT$, $TE$ and $EE$ power spectra to unprecedented precision. In this section we summarize the consequences of the 5-year data for inflation.

\vskip 6pt
\noindent
{\it Flatness of space}

A fundamental prediction of inflation is that the geometry of space should approximate Euclidean flat space to a high degree of accuracy.
According to the recent {\sl WMAP} results\cite{WMAP5} the universe is consistent with
being flat at the 1\% level. 
This implies that the density of the universe is indistinguishable at that level from the critical energy density $\rho_c$ corresponding to a flat universe.
To fulfill this energy budget requires the addition of dark matter and dark energy.
This is of course consistent with the independent evidence for dark matter and dark energy from galaxy surveys and supernova explosions.

\vskip 6pt
\noindent
{\it Superhorizon correlations}

The
{\it WMAP} detection of an anti-correlation between CMB temperature
and polarization fluctuations at angular separations $5^\circ > \theta >
1^\circ$ (corresponding to the {\it TE} anti-correlation seen on
scales $\ell \sim 100-150$ in Fig.~\ref{fig:TE}) is a distinctive
signature of adiabatic fluctuations on superhorizon scales at the
epoch of decoupling, confirming a fundamental prediction of the
inflationary paradigm \cite{TEcorr,spergel/zaldarriaga:1997,peirisetal03}. 
During inflation the superluminal\footnote{It is the scale-factor that evolves superluminally. This is perfectly compatible with General Relativity, and no information is exchanged faster than light.} expansion stretches microscopic quantum fluctuations to macroscopic super-horizon scales.
The observed $TE$ anti-correlation on large scales is remarkable qualitative evidence for this 
basic mechanism.

\begin{table}[btp]
\begin{center}
\begin{tabular}{||c|c|c||}
\hline \hline
\bf Parameter & 5-year {\sl WMAP} & {\sl WMAP}+BAO+SN \\
\hline
 $n_s$ & $0.963_{-0.015}^{+0.014}$ & $0.960_{-0.013}^{+0.014}$ \\
\hline
$n_s$ & $0.986 \pm 0.022$ & $0.968 \pm 0.015$ \\
 $r$ & $<0.43$ & $<0.20$ \\
\hline
 $n_s$ & $1.031_{-0.055}^{+0.054}$ & $1.022_{-0.042}^{+0.043}$ \\
 $\alpha_s$ & $-0.037 \pm 0.028$ & $-0.032_{-0.020}^{+0.021}$ \\
\hline
$n_s$ & $1.087_{-0.073}^{+0.072}$ & $1.093_{-0.069}^{+0.068}$ \\
 $r$ & $<0.58$ & $<0.54$ \\
$\alpha_s$ & $-0.050 \pm 0.034$ & $-0.055 \pm 0.028$ \\
\hline \hline
\end{tabular}
\end{center}
\caption{ \label{tab:param} 5-year {\sl WMAP} constraints on the primordial power spectra in the power law parameterization \cite{WMAP5}.}
\end{table}

\vskip 6pt
\noindent
{\it Measurement of the scalar spectrum}

Komatsu et al.~\cite{WMAP5} recently used the {\sl WMAP} 5-year temperature and polarization
data, combined with the luminosity distance data of Type Ia supernovae (SN) at redshifts $z\le 1.7$ \cite{unionsn} and the
angular diameter distance data of Baryon Acoustic Oscillations (BAO) at redshifts $z=0.2$ and $0.35$ \cite{bao}, to
put constraints on the shape of primordial power spectra (see Fig.~\ref{fig:WMAP5} and Table \ref{tab:param}).

The {\sl WMAP} analysis employed the standard power-law parameterization of the power spectrum\cite{WMAP5}
\beq
\label{equ:Ps1}
P_s(k) = A_s(k_\star) \left( \frac{k}{k_\star}\right)^{n_s(k_\star) -1}\, .
\eeq
The amplitude of scalar fluctuations at $k_\star = 0.002\, {\rm Mpc}^{-1}$ is measured to be
$A_s(k_\star) = (2.445\pm 0.096) \times 10^{-9}$.
From both numerical simulations and analytical estimates one finds that this initial amplitude of density fluctuations gives gravity enough time to form the large-scale structures we observe today.
Assuming {\it no} tensors ($r\equiv 0$) the scale-dependence of the power spectrum is
\beq
n_s = 0.960 \pm 0.013 \quad (r\equiv 0)\, .
\eeq
The scale-invariant Harrison-Zel'dovich-Peebles spectrum, $n_s = 1$, is 2.9 standard deviations away from the mean of the likelihood.

Including the possibility of a non-zero $r$ into the parameter estimation results in the posterior probability in Fig.~\ref{fig:WMAP5}. 
The marginalized constraint on $n_s$ becomes
\beq
n_s = 0.970 \pm 0.015 \quad (r \ne 0)\, .
\eeq
The worsening of the constraint simply means that there is a {\it degeneracy} between $n_s$ and $r$, and the current data cannot simultaneously measure them independently.

\vskip 6pt
\noindent
{\it Constraints on the tensor spectrum}

The data is beginning to put the first meaningful constraints on the tensor spectrum.
The {\sl WMAP} 5-year analysis finds the following upper limit on the tensor-to-scalar ratio
\beq
\label{equ:rlimit}
r < 0.2 \quad (95 \%\, {\rm C.L.})\, .
\eeq
The {\sl WMAP} limits on the $BB$ spectrum are still rather weak, so the result (\ref{equ:rlimit}) is
driven mainly by $TT$ and $TE$ measurements.
Below we discuss the future prospects of CMB polarization experiments for improving the sensitivity to $B$-modes by an order of magnitude or more.

\vskip 6pt
\noindent
{\it Constraints on inflationary models}

\begin{figure}[htbp!]
    \centering
        \includegraphics[width=.45\textwidth]{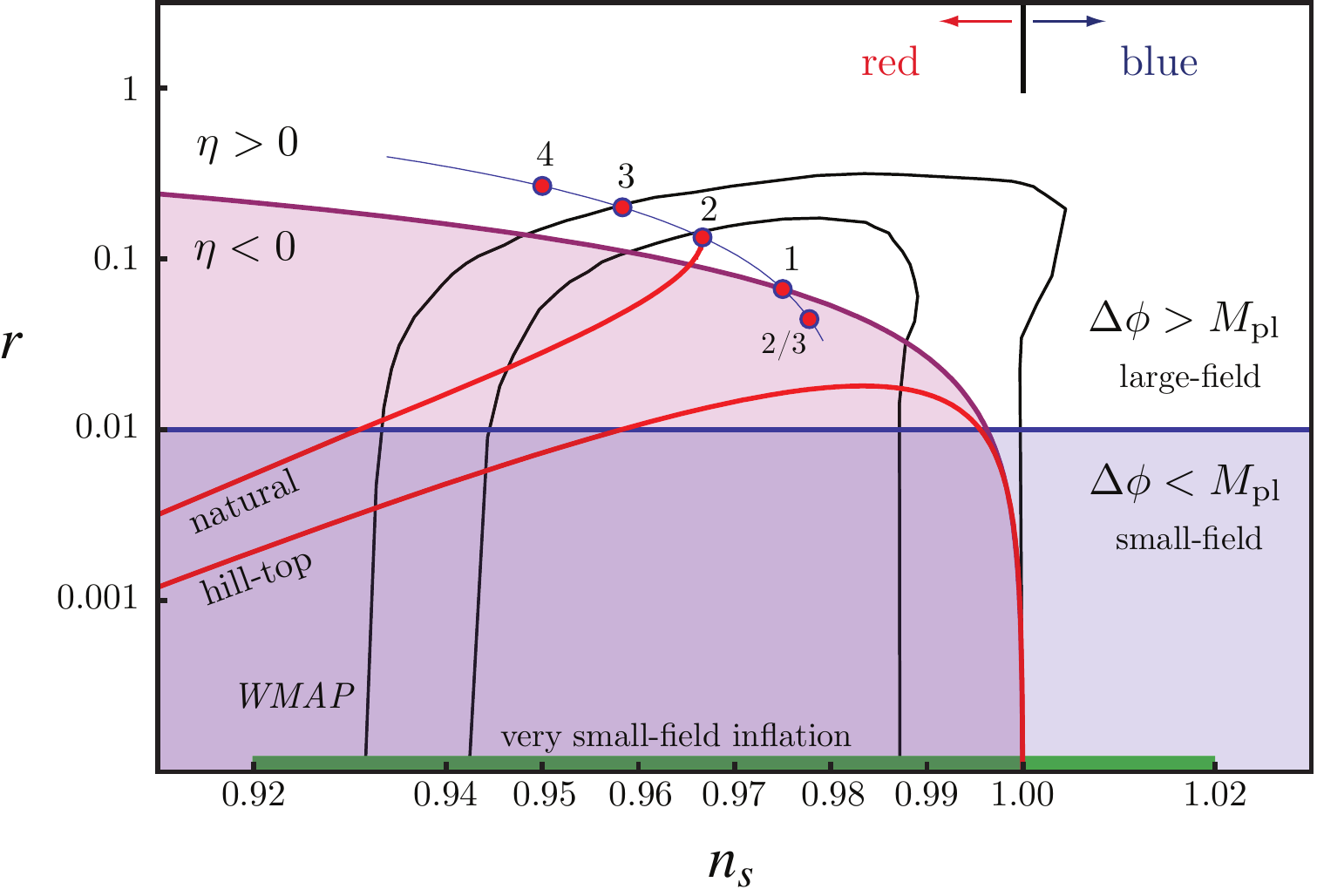}
          \caption{Constraints on single-field slow-roll models in the $n_s$-$r$ plane. The value of $r$ determines whether the model involves large or small field variations.  The value of $n_s$ classifies the scalar spectrum as red or blue. Combinations of the values of $r$ and $n_s$ determine whether the curvature of the potential was positive ($\eta > 0$) or negative ($\eta < 0$) when the observable universe exited the horizon.
   Shown are also the {\sl WMAP} 5-year constraints on $n_s$ and $r$ \cite{WMAP5} as well as
the predictions of a few representative models of single-field slow-roll inflation:
\newline
{\it chaotic inflation}: $\lambda_p\, \phi^p$, for general $p$ (thin solid line) and for $p=4, 3, 2, 1, \frac{2}{3}$({\large $\bullet$}); models with $p=1$~\cite{MSW} and $p=\frac{2}{3}$~\cite{Silverstein:2008sg} have recently been obtained in string theory, 
{\it natural inflation}: $V_0 [1-\cos(\phi/\mu)]$ (solid line), {\it hill-top inflation}: $V_0 [1-(\phi/\mu)^2] + \dots$ (solid line), 
{\it very small-field inflation}: models of inflation with a very small tensor amplitude, $r \ll 10^{-4}$ (green bar); examples of such models in string theory include brane inflation \cite{KKLMMT, delicate1, BDKM, Holographic}, K\"ahler inflation \cite{Conlon:2005jm}, and racetrack inflation \cite{BlancoPillado:2006he}.}
    \label{fig:nsr}
\end{figure}

In this review we intentionally did {\it not} put much emphasis on the potential of the present data to test
inflationary {\it models}.
Personally, we feel that very detailed comparisons between the wide range of inflationary models and the CMB data is somewhat premature.
At the present stage we are still testing basic aspects of the inflationary {\it mechanism} rather than details of its specific implementation.
Having said that, we mention in passing that constraints on specific inflationary models have been obtained (see Fig.~\ref{fig:nsr}):
Inflationary models predicting a blue spectrum ($n_s > 1$) are now virtually ruled out; this includes models of hybrid inflation like $V(\phi) =V_0[1 + m^2 \phi^2]$.
Assuming that the tensor amplitude $r$ is small, the measurement $n_s < 1$ implies a constraint on the curvature of the inflaton potential, $V'' < 0$.
Finally, models predicting a very large tensors amplitude $r$ are ruled out, {\it e.g.} $V(\phi) = \lambda \phi^4$. 
In sum,
many popular models are still allowed by the data, but an increasing number of models are on the verge of being tested seriously.

\section{Future Prospects}
\label{sec:future}

The future of CMB observations is bright.
The European {\sl Planck} satellite\cite{Planck} will measure the temperature power spectrum over a large range of scales with unprecedented accuracy and resolution.
In addition, {\sl Planck} will provide improved constraints on $E$- and $B$-mode polarization.
The {\sl Planck} data will be supplemented by many ground-based or balloon experiments with a special focus on measurements of the small-scale temperature fluctuations and/or the polarization power spectra.
Finally, concrete plans are being made for a next-generation satellite dedicated to the measurement of CMB polarization.
This {\sl CMBPol} experiment proposes to improve the sensitivity to $B$-modes by almost two orders of magnitude.

As we now describe, the combination of this wealth of data will allow detailed tests of the physics of the early universe.

\vskip 6pt
\noindent
{\it Scalars on small scales}

The shape of the spectrum of primordial density fluctuations can be an important diagnostic of the inflationary dynamics.
Equation (\ref{equ:Ps1}) is of course only a simple parameterization of the power spectrum in terms of an amplitude $A_s$ and a spectral index $n_s$, both defined at the pivot scale $k_\star$.
This power law parameterization may be refined by allowing a non-trivial
running of the spectral index $\alpha_s = \frac{d n_s}{d \ln k}$, {\it i.e.} by defining
\beq
P_s(k) = A_s(k_\star) \left(\frac{k}{k_\star} \right)^{n_s(k_\star)-1 + \frac{1}{2} \alpha_s(k_\star) \ln(k/k_\star)}\, .
\eeq
Generically, slow-roll inflation predicts that the running should be a small second-order effect, $|\alpha_s| \sim {\cal O}(0.001)$.  At present, the data isn't good enough to measure $\alpha_s$ values of that order.
In addition, constraints on $n_s$ and $r$ deteriorate if $\alpha_s$ is added to the parameter estimation as an extra parameter (see Table \ref{tab:param}).
However, future small-scale CMB measurements combined with large-scale structure surveys will give a large enough lever arm to provide the first meaningful constraints on $\alpha_s$ or even measure a non-zero value.
In the context of slow-roll inflation it would be hard to explain a value of $\alpha_s$ that is either negative or positive but much larger than 0.001.

\vskip 6pt
\noindent
{\it Tensors}

The {\sl WMAP} upper limit on the amplitude of primordial tensor fluctuations, $r < 0.2$, is only beginning to be a serious constraint on inflationary model-building.
Improved constraints on $B$-modes of CMB polarization will drive future limits on the tensor amplitude (at the moment tensors are still constrained only by $TT$ and $TE$ measurements).
The quest for a $B$-mode detection is one of the most exciting developments in observational cosmology.
The next generation of ground-based and balloon experiments will realistically improve the constraints on tensors to $r\sim {\cal O}(0.01)$.
The proposed {\sl CMBPol} satellite promises improved control over systematic uncertainties and might get down even to $r \sim {\cal O}(0.001)$.
Measuring tensors at that level of sensitivity would mark a qualitative shift in our ability to test the inflationary paradigm.  If tensors aren't seen at that level, all large-field models of inflation are ruled out.
On the other hand,
a $B$-mode detection would be an extraordinary discovery and a ``smoking gun" for inflation.

\vskip 6pt
\noindent
{\it Non-Gaussianity}

For most of this review we have assumed that the primordial fluctuations have a Gaussian distribution.
Indeed, this is a fundamental prediction of slow-roll inflation and the current observational limits confirm this at the 0.1\% level.\footnote{By this measure, non-Gaussianity has been constrained more accurately than curvature.}
However, as we will now discuss, a small degree of primordial non-Gaussianity can be a crucial probe of the inflationary dynamics.

Non-Gaussianity is a measure of {\it inflaton interactions}.
To allow for slow-roll inflation the inflaton field is necessarily weakly interacting ($V(\phi)$ is very flat) and the non-Gaussianity is predicted to be small\cite{malda}.
However, going beyond the single field slow-roll paradigm, both non-trivial kinetic terms (derivative self-interactions in the inflaton action) and the presence of more than one light field during inflation may lead to large, observationally distinct non-Gaussianity.

Gaussian fluctuations are characterized completely by their two-point correlation function,
$\langle \delta \rho \, \delta \rho \rangle$, or equivalently by the power spectrum, $P_s(k)$.
Non-Gaussianity is therefore measured by considering higher-order correlation functions, {\it i.e.}~any result that cannot be explained by the power spectrum alone is a signature of non-Gaussian fluctuations.
The leading effect is given by the three-point function, $\langle \delta \rho \, \delta \rho \, \delta \rho \rangle$, or its Fourier equivalent, the {\it bispectrum}, $B_s(k,k')$.
The bispectrum is measured by sampling {\it triangles} in Fourier space. 
A lot of physical information is contained in the momentum dependence or the shape of the bispectrum.
Studies of non-Gaussianity may become a powerful probe of ultra-high energy physics and inflation.\footnote{The first hints of primordial non-Gaussianity might even already have been detected in the three year\cite{Yadav} and five year\cite{WMAP5} {\sl WMAP} data.}

\section{Conclusions}
\label{sec:conclusions}

Progress in modern cosmology is driven by a 
healthy balance of theory and observations.
In this review we have 
attempted to give a flavor of the excitement felt by many cosmologists and particle physicists in using observations of the cosmic microwave background to learn about the universe at the highest energies and the smallest distance scales.

Inflation, a period of accelerated expansion in the very early universe, explains how regions of space which should be uncorrelated at CMB decoupling are observed to have almost identical temperatures.
In the inflationary paradigm, quantum fluctuations in the very early universe were in fact produced when the relevant scales were causally connected.
Subsequently, however, the superluminal expansion of space during inflation stretched these scales outside of the horizon.
When the perturbations re-entered the horizon at later times, they served as the initial conditions for the growth of large-scale structure and the anisotropies in the CMB.
Inflation makes detailed predictions about key statistical features of the primordial perturbations such as their scale-dependence and (non-)Gaussianity.  In addition, inflation predicts a stochastic background of gravitational waves which leaves a characteristic ($B$-mode) signature in the polarization of the CMB.  If observed, $B$-modes will reveal the energies at which inflation occurred.

Observations of the cosmic microwave background temperature anisotropies and polarization are just becoming precise enough to test the detailed predictions of the inflationary paradigm.
In particular, the data is starting to measure the shape of the primordial perturbation spectra.
The scalar (density) perturbations are found to have almost equal power on all scales.  However, the data also seems to show the characteristic small deviations from perfect scale-invariance that naturally arise from inflation.  
Measurements of density fluctuations alone are not enough to 
be wholly persuasive that inflation indeed occurred since alternative theories for the early universe\cite{ekpyrotic, cyclic} might achieve the same.
However, only inflation could explain a primordial gravitational wave background comparable in magnitude to the density fluctuations.  A future detection of gravitational waves through their effect on CMB polarization would be a revolutionary discovery and a ``smoking gun" for inflation.

Given the 
remarkable recent developments in our theoretical and observational quest to understand the origin, structure and evolution of the universe, we conclude that
{\it the Golden Age of Cosmology is only beginning}.

\acknowledgments HVP is supported in part by Marie Curie grant MIRG-CT-2007-203314 from the European Commission, and by an STFC Advanced Fellowship.
The research of DB is supported in part by the David and Lucile Packard Foundation and the Alfred P.~Sloan Foundation and by Fellowships of the Center for the Fundamental Laws of Nature and the Center for Astrophysics at Harvard. 
HVP gratefully acknowledges the WMAP Science Team, Richard Easther, George Efstathiou, Wayne Hu, and the Cambridge cosmology crew for fruitful collaborations and important insights. 
DB thanks Liam McAllister, Eva Silverstein, Paul Steinhardt and Matias Zaldarriaga for stimulating discussions about physics and cosmology. We acknowledge the use of Figs.~2, 3, 7, 10, and 11 provided by the WMAP Science Team, publicly available on the Legacy Archive for Microwave Background Data (LAMBDA). Support for LAMBDA is provided by the NASA Office of Space Science. We are grateful to Wayne Hu for Fig.~4. 

\vfil

\newpage
\bibliographystyle{AdvSciLett.bst}
\bibliography{InflationReview}

\end{document}